\pdfoutput=1
\documentclass[%
 reprint, superscriptaddress,
 amsmath,amssymb,
 aps, float,
prb,
floatfix
]{revtex4-2}

\usepackage{graphicx, color, soul, tabularx, textcomp, lipsum, gensymb}% Include figure files
 \usepackage[normalem]{ulem}
\usepackage{dcolumn}% Align table columns on decimal point
\usepackage{bm, xcolor, siunitx, float}% bold math
\begin{document}

\preprint{APS/123-QED}

\title{Enhanced Critical Currents and Irreversibility Fields in YBa$_2$Cu$_4$O$_8$ Films through Ca-Substitution}

\author{Jiangteng Liu}
\affiliation{Department of Electrical \& Computer Engineering, University of Washington, Seattle, WA 98195}

\author{Shuhei Funaki}
\affiliation{Faculty of Engineering, Aichi Institute of technology: 1247 Yachigusa Yakusa-chou, Toyota-shi, Aichi 470-0392, Japan}

\author{Yuki Ogimoto}
\affiliation{Graduate School of Science and Technology, Seikei University: 3-3-1 Kichijoji-kitamachi, Musashino-shi, Tokyo 180-8633, Japan}

\author{Bryan Zhang}
\affiliation{Department of Electrical \& Computer Engineering, University of Washington, Seattle, WA 98195}

\author{Ryoya Nagaura}  
\affiliation{Graduate School of Science and Technology, Seikei University: 3-3-1 Kichijoji-kitamachi, Musashino-shi, Tokyo 180-8633, Japan}

\author{Ryuji Yoshida}
\affiliation{Graduate School of Science and Technology, Seikei University: 3-3-1 Kichijoji-kitamachi, Musashino-shi, Tokyo 180-8633, Japan}
\affiliation{Nanostructures Research Laboratory, Japan Fine Ceramics Center:   2-4-1 Mutsuno, Atuta-ku, Nagoya 456-8587, Japan}

\author{Takeharu Kato}
\affiliation{Nanostructures Research Laboratory, Japan Fine Ceramics Center:   2-4-1 Mutsuno, Atuta-ku, Nagoya 456-8587, Japan}

\author{Masashi Miura}
\affiliation{Graduate School of Science and Technology, Seikei University: 3-3-1 Kichijoji-kitamachi, Musashino-shi, Tokyo 180-8633, Japan}
\affiliation{JST-FOREST, 7, Gobancho, Chiyoda-ku, Tokyo 102-0076, Japan}

\author{Serena Eley}
\affiliation{Department of Electrical \& Computer Engineering, University of Washington, Seattle, WA 98195}

\date{\today}

\begin{abstract}

The intrinsic carrier concentration of superconducting materials may not be optimal, and different superconducting properties may have distinct optimal doping levels. Here, we show that Ca substitution increases the hole concentration in intrinsically underdoped epitaxial (Y$_{1-x}$Ca$_x$)Ba$_2$Cu$_4$O$_8$ (YCa124) films with $x \leq 0.10$, driving YCa124 toward optimal doping, increasing $T_c$ and $J_c$. At 40 K, 10\% Ca doping increases $J_c$ by factors of 2.4 and 10 at 0.03 and 6 T, respectively, compared with the undoped film. The field dependence of $J_c$ follows $J_c \propto B^{-0.5}$, independent of Ca content, indicating that Ca substitution does not alter the dominant pinning mechanism. This field dependence is consistent with pinning by planar Y-125 stacking-fault intergrowths observed by electron microscopy. In addition, Ca substitution slows thermally activated vortex motion (creep) at $T \gtrsim 15$ K, while at lower temperatures the creep rate $S$ shows an unusual decrease with increasing field. Finally, we consider how doping tunes the upper bound on $J_c$, set by the depairing current density $J_d$, and the lower bound on creep, set by the Ginzburg parameter $G_i$. We find positive correlations between $J_c$ and $J_d$, and between $S$ and $G_i$, and compare them with trends across a broad range of superconductors.
\end{abstract}

\maketitle

\section{\label{sec:introduction} Introduction}

The superconducting properties of cuprates are strongly governed by the hole concentration in the CuO$_2$ planes, which influences characteristics such as the transition temperature, condensation energy, penetration depth, and vortex dynamics \cite{Deutscher2014,Tallon2001}. Consequently, controlling the carrier concentration provides an important route for understanding and optimizing the critical current density in high-temperature superconductors. In YBa$_2$Cu$_3$O$_{7-\delta}$ (Y123), the carrier density is commonly tuned through oxygen content, where oxygen vacancies in the CuO chains modify the charge transferred to the CuO$_2$ planes \cite{Jorgensen1990,Cava1988}. However, oxygen variation also changes the oxygen ordering and local structural environment of the CuO chains, so effects attributed to carrier concentration are convolved with these structural changes \cite{VonZimmermann2003,PhysRevMaterials.3.114806,Talantsev2014,Gazquez2016}.

An alternative approach is to tune the carrier concentration through Ca substitution. In Y123, partial substitution of trivalent yttrium ions (Y$^{3+}$) with divalent calcium ions (Ca$^{2+}$) increases the hole concentration and has been widely investigated as a strategy for improving supercurrent transport across grain boundaries \cite{Schmehl1999,Daniels2000,Song2005,PhysRevB.64.140508,Weber2003,PhysRevB.51.8582,Yamamoto_2024}. This enhancement has been attributed to an increase in carrier density near the grain boundary, which shortens the electrostatic screening length and reduces the grain-boundary potential barrier, thereby improving the intergranular critical current density \cite{PhysRevLett.92.195502,Song2005}. Accordingly, Ca-doped Y123 films and multilayer architectures incorporating Ca-rich regions are being extensively investigated for improving current transport in rare-earth barium copper oxide (REBCO) conductors \cite{Hammerl2000,Aye2024,Sebastian2023CaDopedYBCO,Ogunjimi2022,PhysRevB.111.205113}.

Beyond its impact on grain-boundary transport, Ca substitution also provides a means of systematically tuning the bulk electronic properties of cuprates through controlled changes in hole concentration. This approach is particularly attractive in YBa$_2$Cu$_4$O$_8$ (Y124), whose stable double CuO chain structure and fixed oxygen stoichiometry enable the carrier concentration to be tuned by Ca substitution across a wide range of hole doping levels, spanning underdoped to overdoped regions of the superconducting phase diagram. As in Y123, Ca$^{2+}$ substitution for Y$^{3+}$ in Y124 increases the hole concentration in the CuO$_2$ planes, but here it drives the material from an intrinsically underdoped state toward optimal doping and, at sufficiently high Ca concentrations, into the overdoped regime \cite{Miyatake1989,Berastegui1994, SCHWER199496, Fischer1993}. 

Previous studies of Ca-substituted Y124 powders and polycrystalline samples have focused on synthesis, structural characterization, as well as the evolution of the superconducting transition temperature and superfluid density with doping \cite{Miyatake1989,Berastegui1994,Yang2008,PhysRevB.58.3457}. Here, we present a systematic study of how Ca-induced changes in hole concentration influence the critical current density and vortex dynamics in epitaxial (Y$_{1-x}$Ca$_x$)Ba$_2$Cu$_4$O$_8$ (YCa124) films. Using magnetization measurements, we determine the critical current density $J_c(T,H)$ across a range of temperatures and magnetic fields, as well as the rate of thermally activated vortex motion $S(T,H)$. Scanning transmission electron microscopy is employed to characterize the film microstructure. Finally, we compare these results with those reported for other superconducting materials.

\section{Results and Discussion}

(Y$_{1-x}$Ca$_x$)Ba$_2$Cu$_4$O$_8$ films with five Ca concentrations ($x = 0$, 0.025, 0.05, 0.075, and 0.10) were grown via a molten hydroxide method, described in the Methods section, on NdGaO$_3$ (001) substrates. Before calculating hole concentrations for our films, we consider how reliably the nominal Ca content $x$ tracks the actual hole doping level. The relationship between Ca substitution and hole concentration in YCa124 has been examined using nuclear quadrupole resonance (NQR) and neutron diffraction. Calcium substitutes predominantly on the Y site, consistent with divalent Ca$^{2+}$ replacing trivalent Y$^{3+}$ increasing the hole concentration in the CuO$_2$ planes \cite{SCHWER199496, Fischer1993}. Machi et al. supported this picture, reporting a suppressed antiferromagnetic spin-fluctuation contribution to the $^{63}$Cu spin-lattice relaxation rate with Ca doping, consistent with increased hole concentration \cite{MACHI1992635}.  At higher doping levels, Mali et al. found that the effective Y-site substitution fraction $x'$ falls somewhat below the nominal Ca content $x$, attributed to a growing contribution from Ba-site substitution \cite{PhysRevB.53.3550}. Because Ca$^{2+}$ substituting for the isovalent Ba$^{2+}$ adds no holes, the growing shift toward Ba-site occupancy at high doping would cause the true hole concentration to fall below what the nominal Ca content $x$ would suggest. Given this residual uncertainty at high $x$, we use our measured critical temperature $T_c(x)$ to estimate the hole concentration $p$ in our films rather than assuming a fixed relation of $\Delta p(x) \approx x/2$ per CuO$_2$ plane.

\begin{figure}[!ht]
\centering
\includegraphics[width=1\linewidth]{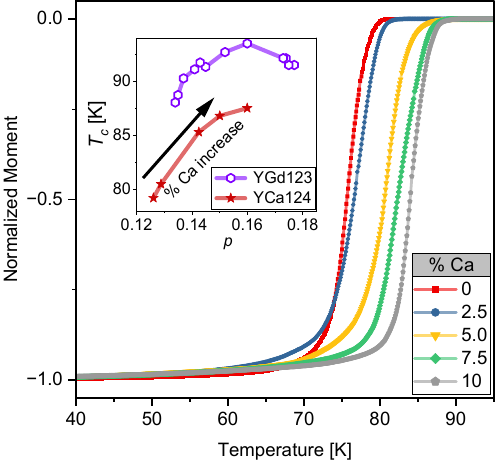}
\caption{\label{fig:fig1} Normalized magnetic moment as a function of temperature for all 5 samples in an applied field of 3 Oe. The inset shows the hole doping dependence of $T_c$. YGd123 data are from Ref. \cite{Miura2022}.}
\end{figure}

To extract the critical temperature $T_c$, we performed magnetization studies. All magnetization measurements in this study were collected using a Quantum Design MPMS3 superconducting quantum interference device (SQUID) magnetometer. Figure \ref{fig:fig1} shows the superconducting transition for each film, obtained from measurements of the magnetic moment collected in a small applied field of 3 Oe after zero-field cooling (ZFC). Using the empirical relation $T_c/T_{c,max} = 1 - 82.6(p-0.16)^2$ \cite{Presland1991} we convert our measured $T_c(x)$ into the corresponding hole concentration $p(x)$, and plot $T_c$ versus $p$ in the inset to Fig. \ref{fig:fig1}. These results confirm that Y124 is intrinsically underdoped: $T_c$ increases with Ca doping across our full range, reaching its highest value at $x=0.1$, the maximum doping level studied. Higher Ca content may be needed to reach optimal doping. For comparison, the characteristic superconducting dome of $T_c$ vs.\ $p$ for (Y,Gd)Ba$_2$Cu$_3$O$_y$ (YGd123) is also shown, using data from Ref.~\cite{Miura2022}.

Cuprate superconductors are predicted to have among the highest intrinsic upper limits on the critical current density $J_c$ of any known class of superconductors. This limit is set by the depairing current density $J_d$, the current at which Cooper pairs break because the kinetic energy of the supercurrent exceeds the condensation energy, which scales with the energy gap\cite{Tinkham1996, DewHughes2001, Blatter1994, Miura2022, Ruiz2026, Eley2021}.  In practice, however, several mechanisms suppress the attainable $J_c$ well below $J_d$. A primary limitation is dissipation arising from vortex motion: in applied magnetic fields, quantized magnetic flux lines (vortices) penetrate the material, and their motion dissipates energy and degrades $J_c$ \cite{PhysRevB.41.8986, MacManus-Driscoll2021, Kwok2016, Puig2024, Eley2017, Eley2017, Ruiz2026, Liu2026}. Defects can pin these vortices and suppress their motion, so vortex-defect interactions are a key determinant of $J_c$. Tuning the carrier concentration, which governs both $J_d$ and the thermodynamic critical field $H_c$, has also been shown to be an effective route for enhancing $J_c$ \cite{Miura2022, Miura2024, Kethamkuzhi2026}.

\begin{table}[h!] 
\caption{\label{tab:samples} \textbf{YCa124 sample characteristics}: Ca content ($x$, \%), doping level ($p$), $T_c$ (K), coherence length $\xi_{ab}$ (nm), penetration depth $\lambda_{ab}$ (nm), upper critical field $H_{c2}$ (T), irreversibility field $H_{\mathrm{irr}}$ (T), and depairing current density $J_d$ (MA/cm$^2$), all at $T=0\ $ K; $H_{\mathrm{irr}}$ is at $T/T_c=0.6$ and anisotropy $\gamma$ is at $T/T_{c2}(0 \ \mathrm{T})=0.96$. The Ginzburg number $G_i$ is calculated using the listed parameters. }
\begin{tabularx}{1\linewidth}{cllcccccccc}
\hline \hline
    ID & Ca & $p$ & $T_c$ & $\xi_{ab}$ & $\lambda_{ab}$ & $H_{c2}$ & $H_{irr}$ & $\gamma$ & $J_d$ & $ G_i\times 10^3$ \\
    \hline
    1 & $0$ & $0.126$ & $79.2$ & $4.30$ & $157.0$ & $17.8$ & $3.4$ & $8.64$ & $95.1$ & $5.000$ \\
    2 & $2.5$ & $0.129$ & $80.5$ & $3.44$ & $155.9$ & $27.8$ & $3.3$ & $8.98$ & $120.6$ & $8.477$ \\
    3 & $5.0$ & $0.143$ & $85.3$ & $3.40$ & $154.9$ & $28.4$ & $4.5$ & $8.53$ & $123.6$ & $8.568$ \\
    4 & $7.5$ & $0.150$ & $86.8$ & $2.84$ & $153.9$ & $40.7$ & $5.4$ & $5.89$ & $149.9$ & $5.908$\\
    5 & $10$ & $0.160$ & $87.5$ & $2.46$ & $152.8$ & $54.2$ & $5.5$ & $4.42$ & $175.5$ & $4.379$\\
\hline \hline
\end{tabularx}
\end{table}

\begin{figure*}[t!]
\centering
\includegraphics[width=1\linewidth]{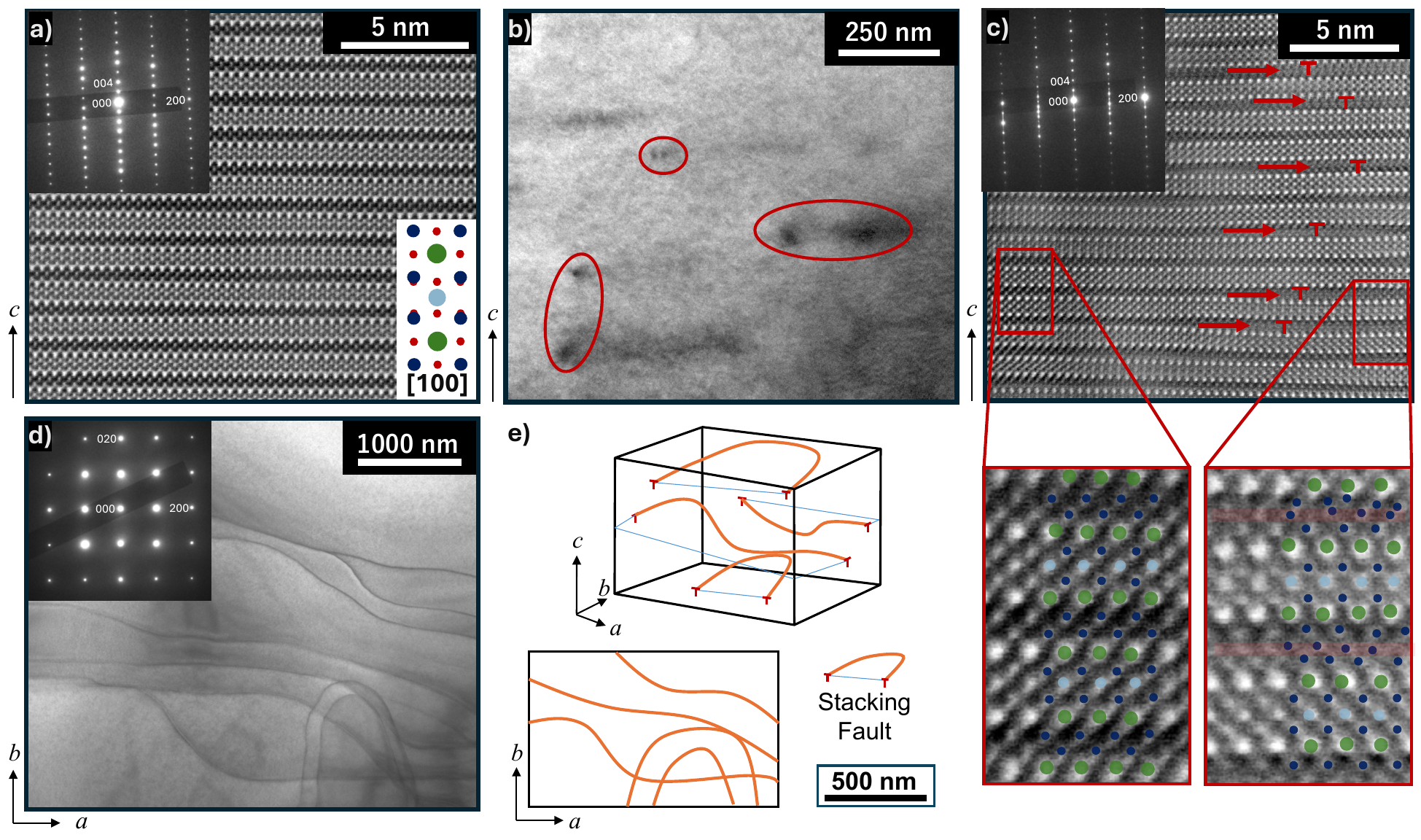}
\caption{\label{fig:fig2} (a) Cross-sectional HAADF-STEM images of sample with 7.5\% Ca doping. The upper and lower insets show the electron diffraction pattern and crystal structure of YBa$_2$Cu$_4$O$_8$ viewed along the [100] direction, respectively. The atom colors are Y/Ca (pale blue), Ba (Green), Cu (dark blue), and O (red). Cross-sectional view of (b) TEM images and (c) high-angle annular dark-field STEM images of the YCa124 film in the stacking fault region, where the electron diffraction pattern also shows extra diffraction spots. The red arrow indicates the presence of stacking faults caused by extra CuO layers. This is also highlighted in red in the zoomed-in view. (d) Planar-view TEM image showing the stacking fault. The upper insets show the electron diffraction pattern. (e) Illustrations that reveal the microstructure information for the YCa124 film. All images are taken from the sample with 7.5\% Ca content.}
\end{figure*}

Within the Ginzburg-Landau framework, $J_d$ is given by
\begin{equation}
J_d = \frac{2\sqrt{2}}{3\sqrt{3}}\frac{H_c}{\lambda_{ab}}=\frac{\Phi_0}{3^{3/2}\pi \mu_0 \lambda_{ab}^2\xi_{ab}}
\end{equation}
where $H_c = \Phi_0/(2\sqrt{2}\pi\mu_0\lambda_{ab}\xi_{ab})$ is the thermodynamic critical field, $\lambda_{ab}$ is the penetration depth, and $\Phi_0$ is the magnetic flux quantum. Increasing $H_c$ reduces thermal fluctuations, since the Ginzburg number scales as $G_i \propto T_c^2\gamma^2H_c^{-4}\xi_{ab}^{-6}$. The Ginzburg number quantifies the strength of thermal fluctuations relative to the condensation energy. A larger $G_i$ means thermal fluctuations more readily disrupt vortex order, for example melting the vortex lattice into a liquid well below $H_{c2}$, as seen in the cuprates' wide vortex-liquid regions compared to conventional low-$T_c$ superconductors \cite{Blatter1994}. The Ginzburg number is also predicted to set a lower limit on the thermally induced vortex creep parameter $S$, through $S_{\min}\sim G_i^{1/2}(T/T_c)$ in the Anderson-Kim regime (low temperatures, single vortex dynamics) \cite{Eley2017}. 

To probe how Ca substitution tunes the intrinsic $J_d$ ceiling and the strength of thermal fluctuations, we calculate $J_d$ and $G_i$ from the doping dependence of $\xi_{ab}$, $\lambda_{ab}$, $\gamma$, and $T_c$, summarized in Table~\ref{tab:samples}. Penetration depth values are from Refs.~\cite{Khasanov2004, Shengelaya1998}. Coherence length and anisotropy measurements are detailed in Supplementary Fig.~S1--S4.

To better understand extrinsic factors that affect $J_c$, such as vortex-defect interactions, we characterize the defect landscape via transmission electron microscopy. To interpret the structure revealed in the images, it is useful to note that Y124 can be understood as Y123 with an extra CuO row inserted between the BaO layers. This addition converts the single rows of corner-sharing CuO$_4$ square groups found in Y123 into double rows of edge-sharing CuO$_4$ groups, giving Y124 its characteristic double Cu-O chains \cite{Domenges1991}.

\begin{figure*}[t!]
\centering
\includegraphics[width=1\linewidth]{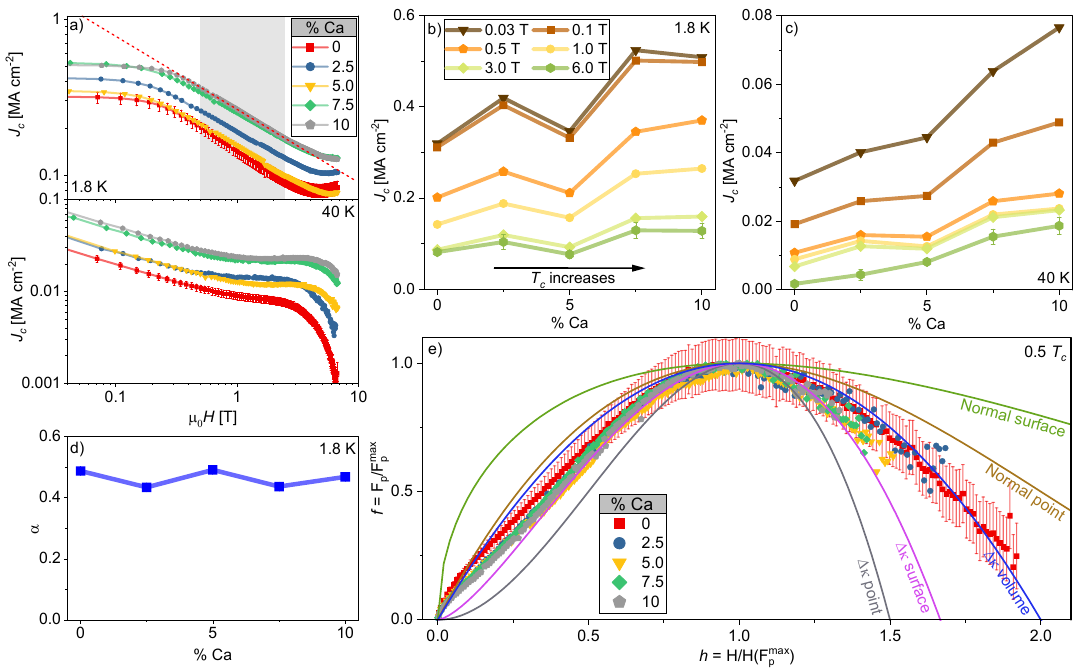}
\caption{\label{fig:fig3} (a) Field-dependent $J_c$ for YCa124 films with different Ca doping levels at $T=1.8 \text{ K}$ in the upper panel and 40 K in the lower panel. Critical current density $J_c$ versus \% Ca at different applied magnetic fields for temperatures of (b) 1.8 K and (c) 40 K, respectively. (d) Comparison of extracted $\alpha$ vs \% Ca at 1.8 K. Here, \(\alpha\) is the power-law exponent extracted from the slope of a linear fit to $\log J_c - \log \mu_0H$, restricted to a field range of 0.5 - 2.5 T, as shown in the shaded region in (a). (e) Normalized pinning force $f=F_p/F_{p,max}$ against reduced field $h=H/H(F_{p, max})$ at $0.5 \ T_c$ for all five YCa124 films. }
\end{figure*}

Figure \ref{fig:fig2}(a) displays a high-angle annular dark field scanning transmission electron microscopy (HAADF-STEM) image of sample 4 with a Ca content of $x=0.075$, revealing Y124's layered structure. Y124 crystallizes into an orthorhombic crystalline structure belonging to the Ammm space group. Figure \ref{fig:fig2}(a) was collected from a particularly well-ordered region of the film, whereas the electron microscopy images in Fig. \ref{fig:fig2}(b-d) are more representative of the sample as a whole,  presenting a multi-scale imaging layout that highlights the spatial geometry and crystallographic nature of the defect landscape. Fig. \ref{fig:fig2}(b) provides a lower-magnification, cross-sectional overview via transmission electron microscopy (TEM) with the crystal \textit{c}-axis oriented vertically. Along the horizontal \textit{ab} planes, distinct dark segments, with examples circled, are distributed aperiodically throughout the matrix. These segments correspond to edge-on terminations of planar defects, visible as local changes in secondary-electron emission.

Figure \ref{fig:fig2}(c) resolves the atomic-scale origin of these dark features through cross-sectional HAADF-STEM, with the \textit{c} axis again vertically oriented. The red arrows indicate local expansion along the \textit{c} axis. To investigate the origin of this expansion, enlarged views of both well-ordered and expanded regions are shown. In the enlarged view of the well-ordered region on the right, we see superconducting CuO$_2$ planes (bright white layers in the STEM image) lying parallel to the \textit{ab} plane, with double Cu-O chains (darker layers) interleaved between the CuO$_2$ bilayers. In contrast, the expanded region on the left contains an additional Cu-O chain layer within the double-chain block, forming a triple-chain segment. This interpretation is also consistent with the additional diffraction spot in the selected-area electron diffraction (SAED) inset, which is absent from the SAED pattern of the well-ordered region shown in Figure \ref{fig:fig2}(a). Therefore, we attribute these local expansions to aperiodic stacking fault segments. In fact, a study has shown these can be associated with intergrowths of the triple-chain YBa$_2$Cu$_5$O$_x$ (Y-125) phase \cite{Zhang2018}.

To examine the morphology of these intergrowths within the \textit{ab} plane, we collected a low-magnification planar-view TEM image, shown in Fig. \ref{fig:fig2}(d). Here, the defect boundaries appear as sweeping arcs of dark contrast, outlining isolated, two-dimensional Y-125 intergrowth domains bounded by curved partial dislocation loops. An additional planar-view STEM image is included in Supplementary Fig.~S5. Lastly, Fig. \ref{fig:fig2}(e) presents 3D and 2D schematics illustrating the spatial configuration of the stacking faults within the crystal.

Ca$^{2+}$ substitution for Y$^{3+}$, being aliovalent, introduces local cationic disorder and structural distortion that appears to favor the formation of Y-125 stacking faults, analogous to the well-documented formation of Y124-type stacking faults in Ca-doped and oxygen-varied Y123 \cite{Talantsev2013}. These short, highly curved segments have dislocation boundaries that bow within the \textit{ab} plane to minimize line energy around local strain and chemical variations. Because these distortions occur on the scale of the superconducting coherence length $\xi_{ab}$, they may act as efficient, isotropic flux pinning centers that restrict vortex motion and enhance the in-field critical current density.

\subsection*{Critical current density}

To determine the temperature and field dependence of the critical current density in our YCa124 films, we carried out magnetization measurements, recording magnetic hysteresis loops at several fixed temperatures and converting the resulting $m(T,H)$ data into $J_c(T,H)$ using the Bean critical state model for rectangular samples \cite{Gyorgy1989, Talantsev2024BeanModel}. Specifically, we applied $J_c(T,H) = 20\Delta m(T,H)/w^2l\delta(1-w/3l)$, where $\Delta m$ is the separation between the upper and lower branches of the hysteresis loop at a given field, and $w$, $l$, and $\delta$ are the film width, length, and thickness, respectively. Here, $J_c$, $m$, the sample dimensions, and the coefficient 20 carry units of A cm$^{-2}$, emu, cm, and A cm$^{-2}$ emu$^{-1}$, respectively \cite{Talantsev2024BeanModel}. Supplementary Fig.~S6 presents the full set of magnetization loops, and further measurement details are given in the Methods section.

With $J_c$ extracted for all films, we next examine how the critical current density varies with field, temperature, and Ca doping. Figure~\ref{fig:fig3}(a-c) compares the field dependence of the critical current density for our films at temperatures of 1.8~K (panel a) and 40~K (panel b). At 1.8~K and low magnetic fields, $J_c$ is approximately field-independent up to a characteristic field $\mu_0H_{sf}$, defining the self-field regime \cite{Polat2011, Abou_El_Hassan2021}. Beyond $\mu_0H_{sf}$, $J_c$ transitions into a distinct field-dependent regime. Above $\mu_0H_{sf}$ and up to $\sim 4$~T at 1.8~K, and below $\sim 1$~T at 40~K, $J_c$ follows a power-law dependence, $J_c\propto B^{-\alpha}$. At higher fields, $J_c(T)$ flattens, even exhibiting a slight increase, corresponding to a weak second magnetization peak (SMP) in $m(H)$. In the 40~K data, we see that as the applied field approaches the irreversibility field, $J_c$ decreases more rapidly.

Having established the field dependence of $J_c$, we next focus on the impact of Ca doping. Figure~\ref{fig:fig3}(b,c) highlights how $J_c$ increases as a function of Ca doping, with a slight dip in the trend for the $x=0.05$ film. Previous studies have found that a doping-induced increase in $J_d$ (and $H_c$) correlates with an increase in $J_c$\cite{Miura2024}. From the approximations in Table~\ref{tab:samples}, $J_d$ increases by a factor of 1.27--1.85 between Ca doping levels of 2.5\% and 10\%. This is consistent with the increase in $J_c(1.8\text{ K}, 0.03\text{ T})$ that we observe, from a factor of 1.08--1.59 over the same doping range. The enhancement in $J_c$ is more prominent at higher magnetic fields; for example, at 40 K, Ca-doping increases $J_c$ by a factor of 2.41 at 30 mT versus 10.18 at 6 T.  
This raises the question of whether the enhancement in $J_c$ reflects a change in the vortex pinning mechanism itself, or simply a strengthening of an unchanged pinning mechanism, driven by the higher carrier concentration that Ca doping introduces. To distinguish between these possibilities, we consider the power-law dependence $J_c \propto H^{-\alpha}$ common to both weak collective pinning \cite{Larkin1979} and strong pinning \cite{Willa2017, EleyStrongpinning2021}. In the former, flux lines are pinned by the collective action of many atomic-scale defects; in the latter, they are pinned by individual larger defects. The value of the power-law exponent $\alpha$ provides insight into the dominant mechanism and the type of defect responsible for pinning. 

We extract $\alpha$ from $J_c(B)$ at 1.8~K for all films and plot it in Fig.~\ref{fig:fig3}(d) versus Ca concentration. We find $\alpha \approx 0.5$ across all films, indicative of vortex pinning by dislocations \cite{Diaz1998, Gurevich1994}, consistent with stacking faults and dislocation loops being the dominant observable defects in our microscopy studies. Ca doping therefore does not change the pinning mechanism, but strengthens pinning at these existing dislocation sites, consistent with the doping-induced increase in $J_d$ and $H_c$ discussed above.

Next, we analyze the pinning force $F_p = \mu_0 H J_c$~\cite{Kramer1973,Hughes1974,Yamasaki1993,Koblischka1998,Yang2008,Yamamoto2009,Fang2011,Luo2025} in these films. In Fig.~\ref{fig:fig3}(e), we compare the normalized pinning force $f = F_p/F_{p,\mathrm{max}}$ against the reduced field $h = H/H(F_{p,\mathrm{max}})$, where $F_{p,\mathrm{max}}$ is the maximum pinning force and $H(F_{p,\mathrm{max}})$ is the field at which this maximum occurs. The curves for all films overlap, consistent with the doping-independent $\alpha$ found above and further indicating that Ca doping does not change the pinning mechanism. To identify the dominant pinning mechanism, we compare the data to five pinning scenarios described by the Dew-Hughes model \cite{Hughes1974, Luo2025, JLiu2026SUST}:
\begin{align}
    \Delta\kappa \text{ point pinning:} \quad & f = 3 h^2 \left(1 - \frac{2h}{3}\right) \label{eq:dk}\\
    \Delta\kappa \text{ surface pinning:} \quad & f = \frac{5}{2} h^{3/2} \left(1 - \frac{3h}{5}\right) \label{eq:dksurf}\\
    \Delta\kappa \text{ volume pinning:} \quad & f = 2 h \left(1 - \frac{h}{2}\right) \label{eq:dkvol}\\
    \text{Normal point pinning:} \quad & f = \frac{9}{4} h \left(1 - \frac{h}{3}\right)^2 \label{eq:normpin}\\
    \text{Surface pinning:} \quad & f = \frac{25}{16} h^{1/2} \left(1 - \frac{h}{5}\right)^2 \label{eq:surfpin}
\end{align}

These five pinning cases are distinguished by both the superconducting properties and the geometry of the pinning centers. $\Delta\kappa$ pinning occurs when defects locally alter the Ginzburg--Landau parameter $\kappa = \lambda/\xi$ (with $\lambda$ the penetration depth and $\xi$ the coherence length), creating spatial variations in vortex free energy. The point, surface, and volume cases are distinguished by the dimensions of the regions with modified $\kappa$ relative to the intervortex spacing $d$. For $\Delta\kappa$ point pinning, these regions are smaller than $d$ in all three dimensions. For $\Delta\kappa$ surface pinning, they extend beyond $d$ in two dimensions, as can occur at grain boundaries or dislocation arrays. For $\Delta\kappa$ volume pinning, they exceed $d$ in all three dimensions and accommodate multiple vortices, with pinning arising at the interfaces between these regions and the surrounding matrix \cite{Hughes1974}. Normal point pinning arises from localized non-superconducting inclusions or defects that suppress superconductivity, lowering the energy cost of a vortex core overlapping the defect. Normal surface pinning involves the same type of energy reduction at extended planar regions where superconductivity is suppressed. Thus, planar features such as grain boundaries, stacking faults, and plate-like precipitates may act as $\Delta\kappa$ or normal surface pins, depending on their local superconducting properties \cite{Hughes1974}.

The solid curves in Fig.~\ref{fig:fig3}(e) show the Dew-Hughes pinning models described by Eqs.~(\ref{eq:dk})--(\ref{eq:surfpin}). We can see that, for $h<1$, pinning in all five films lies between the $\Delta\kappa$ surface and volume pinning predictions. This behavior suggests pinning associated with spatial variations in the superconducting parameters, consistent with the planar Y-125 intergrowths and associated partial dislocation loops observed in Fig.~\ref{fig:fig2}. Such triple-chain intergrowths have also been reported in superoxygenated YBCO films \cite{Zhang2018} and may modify the local superconducting properties through changes in chain structure and strain. For $h>1$, the films with 0 and 2.5\% Ca approximately follow the $\Delta\kappa$ volume pinning curve, whereas increasing Ca content shifts the curves toward the $\Delta\kappa$ surface pinning prediction. This evolution suggests an increasing contribution from planar defects, potentially reflecting a higher stacking fault density with Ca substitution. Field-dependent $J_C$ at other temperatures are included in Supplementary Fig.~S7.

\subsection*{Vortex creep}
\begin{figure*}[t!]
\centering
\includegraphics[width=0.8\linewidth]{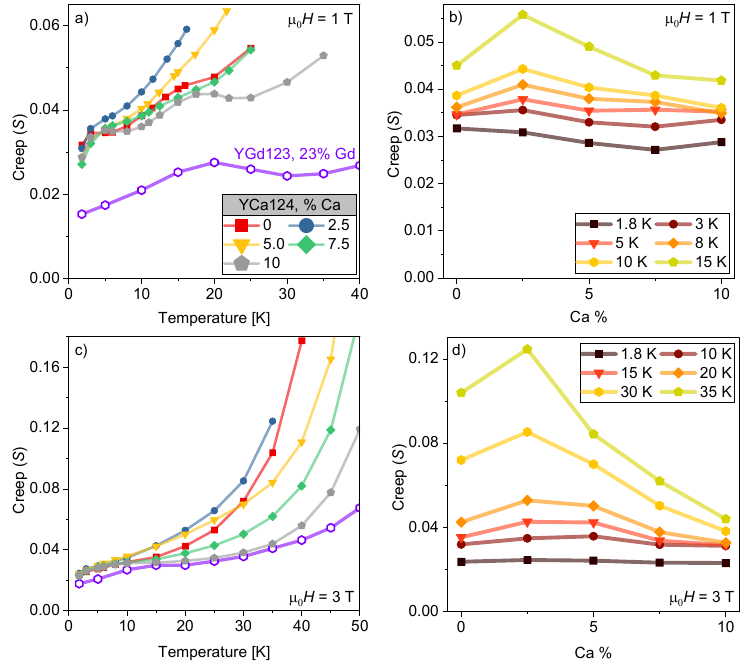}
\caption{\label{fig:fig4}
Temperature dependence of the vortex creep parameter $S$ in YCa124 films with different Ca concentrations under applied magnetic fields of (a) 1 T and (c) 3 T. For comparison, data for a YGd123 film from Ref.~\cite{EleyStrongpinning2021} are included as open hexagons. $S$ versus Ca doping at six different temperatures under an applied magnetic field of (b) 1 T and (d) 3 T.
}
\end{figure*}

\begin{figure*}[t!]
\centering
\includegraphics[width=1\linewidth]{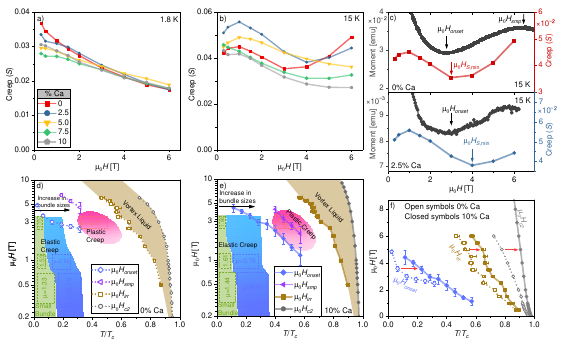}
\caption{\label{fig:fig5} Field dependence of vortex creep parameter $S$ in YCa124 films with different Ca concentrations at (a) 1.8 K and (b) 15 K. (c) Field dependence of $S$ (red and gray symbols; right axis) plotted together with the upper branch of magnetic hysteresis loops (purple curves; left axis) measured at 15 K. The upper and lower panels correspond to films with 0\% and 2.5\% Ca, respectively. The second magnetization peak field $H_{\mathrm{smp}}$, its onset field $H_{\mathrm{onset}}$, and the field where creep is minimum $H_{S,\min}$ are indicated by arrows. Vortex phase diagrams for films with (d) 0\% Ca and (e) 10\% Ca. (f) Effect of the Ca addition on $H_{\mathrm{onset}}$, the irreversibility field $H_{\mathrm{irr}}$, and upper critical field $H_{\mathrm{c2}}$ at different temperatures. }
\end{figure*}

Beyond $T_c$ and $J_c$, vortex creep serves as a sensitive probe of the pinning landscape and how it evolves with Ca doping, since creep is governed by the pinning energy scale $U_0$. Ca substitution affects vortex pinning through two distinct mechanisms, one electronic and one structural. First, Ca doping changes the carrier concentration $n_s$, and because the penetration depth scales as $\lambda \propto n_s^{-1/2}$, this can alter the spatial extent of the supercurrents encircling the vortex core, which extend over a distance of order $\lambda_{ab}$. Second, the ionic size mismatch between Ca$^{2+}$ ($r=1.12$~\AA) and Y$^{3+}$ ($r=1.02$~\AA) expands the \textit{c}-axis spacing between adjacent oxygen planes, directly altering the pinning landscape \cite{SCHWER199496}. Together, these electronic and structural changes provide two routes by which Ca doping can modify vortex pinning and dynamics.

To understand how these changes manifest in vortex dynamics, we next consider the thermally activated motion of the vortex lattice.  A lattice of magnetic flux lines (vortices) forms in type-II superconductors exposed to magnetic fields between the lower, $H_{c1}$, and upper, $H_{c2}$, critical fields. The lattice structure and dynamics depend on the competition between current-driven forces propelling vortices, material disorder that pins them in energy wells $U(J)$, thermal activation out of pinning sites, and elastic deformation of the vortex system. Vortex creep, the thermally activated hopping of vortices between pinning sites under an applied current, therefore provides a direct probe of the pinning energy barriers, vortex-lattice structure, and dynamical behavior. Material defects and disorder create potential wells that lower the vortex free energy by $U_0$ and impede motion. An applied current tilts this landscape, reducing the activation barrier to $U(J)$, while thermal fluctuations allow vortices to escape, causing the current density to decay from $J_{c0}$ to $J(t)$ through magnetic relaxation. Within collective creep theory, the current-dependent activation barrier is \cite{Malozemoff1991, Blatter1994, Yeshurun1996}
\begin{align}\label{eq:Uinterpolation}
U(J)=U_0 [(J_{c0}/J)^\mu -1] / \mu,
\end{align}
\noindent where the glassy exponent $\mu$ characterizes the size of the vortex bundle involved in the creep process.  Notably, $\mu$ depends on whether the object that hops is a single vortex or a vortex bundle of lateral dimension smaller than (small bundle), comparable to (medium bundle), or larger than (large bundle) the penetration depth $\lambda_{ab}$.  Specifically, $\mu$ is predicted to be $1/7$ for the small bundle regime, $3/2$ or $5/2$ for the small bundle regime, and $7/9$ for the large bundle regime \cite{Blatter1994, VINOKUR1995179}.

This activation barrier determines the rate of thermally activated vortex motion. Specifically, the time required for thermal activation over this barrier follows the Arrhenius relation $t=t_0 e^{U(J)/k_BT}$, where $t_0$ is the microscopic time scale associated with the attempt frequency. Typical values are $t_0\sim10^{-8}-10^{-6}$~s \cite{Blatter1994, blatter_vortex_2003, Kwok2016}, although an accurate estimate can also depend on the electric field generated during the magnetic-field sweep \cite{Gurevich1993a, Gurevich1994b}. Combining this relation with Eq.~\ref{eq:Uinterpolation} gives the standard expressions for the time-dependent current and creep rate \cite{Blatter1994, PhysRevB.42.6784, Malozemoff1991, Yeshurun1996}:
\begin{align}\label{eq:Jtdecay}
J(t) \propto M(t)=M_{0} \Big[1+\frac{\mu k_B T}{U_0} \ln (t/t_0)\Big]^{-1/\mu}
\end{align}
\begin{align}\label{eq:ST}
S \equiv \left| \frac{d \ln M}{d \ln t} \right| = \frac{k_B T}{U_0+\mu k_B T \ln (t/t_0)}.
\end{align}
Because $M(t)\propto J(t)$, magnetic-relaxation measurements of $M(t)$ allow extraction of $S$ from $\log M$--$\log t$ plots using Eq.~(\ref{eq:ST}). More details of the measurement and extraction procedure are provided in the Methods section.

With this framework established, we first consider the temperature dependence of the measured creep parameter. Figure~\ref{fig:fig4}(a,c) compares $S(T)$ for our samples at magnetic fields of 1~T and 3~T, respectively. At low temperatures, $T\lesssim10$~K, the $S(T)$ curves overlap and then separate with increasing temperature. Although $S(T)$ generally increases monotonically with temperature, a feature near 18~K may indicate a change in the dominant vortex dynamics, as discussed below.

The emergence of this separation with increasing temperature reveals a systematic dependence on Ca doping. Throughout the measurement range, the sample with 2.5\% Ca exhibits the fastest creep, whereas the sample with 10\% Ca exhibits the slowest. Figure.~\ref{fig:fig5}(b,d) provides a direct comparison of $S$ versus Ca concentration, showing that $S$ is relatively insensitive to doping at low temperatures ($T\lesssim5$~K at 1~T and $T\lesssim10$~K at 3~T), but becomes more strongly doping dependent at higher temperatures. Together with the $J_c$ results, these measurements indicate that 10\% Ca doping simultaneously maximizes $J_c$ and minimizes vortex creep. Additional $S(T)$ at 0.5~T is included in Supplementary Fig.~S8.  

We next compare vortex creep in YCa124 with that in YGd123 without artificial pinning centers from Ref.~\cite{EleyStrongpinning2021}, shown by the purple open hexagons in Figs.~\ref{fig:fig4}(a,c). At 1~T, YGd123 exhibits substantially slower vortex creep than YCa124. For example, at 15~K, the creep rate in YGd123 is only about 60\% of that in the 10\% Ca YCa124 film. Interestingly, this difference becomes much smaller at higher magnetic fields. At 3~T, the two systems exhibit comparable creep rates over the temperature range of 15--35~K, with the creep rate of 10\% Ca YCa124 at 15~K being only about 5\% higher than that of YGd123. This convergence suggests a stronger field dependence of vortex creep in YCa124 and motivates a more detailed examination of $S(H)$.

Figure \ref{fig:fig5} presents the field-dependent $S$ and corresponding vortex phase diagrams for the different Ca concentrations. Panels (a) and (b) show the field dependence of $S$ at 1.8~K and 15~K, while panel (c) compares $S(H)$ with the corresponding $M(H)$ measurements. Panels (d) and (e) show the resulting $H$--$T$ vortex phase diagrams, and panel (f) highlights the main changes in the phase diagram with Ca doping. First focusing on $S(H)$, at 1.8~K, $S$ monotonically decreases with field for all samples, an unexpected trend since increasing vortex density often increases $S$, as observed in e.g., BaFe$_2$(A$_{1-x}$P$_x$)$_2$ \cite{Eley2017}, (Ba/Sr)$_{0.6}$K$_{0.4}$Fe$_2$As$_2$ \cite{Dong2019}, NbSe$_2$ \cite{Eley2018}, FeSe \cite{Lanoel2024}, KCa$_2$Fe$_4$As$_4$F$_2$ \cite{Pyon2021}, YBCO \cite{Yeshurun1996}, and MgB$_2$ \cite{PhysRevB.64.134505}. For the 0\% Ca sample, for example, $S$ at 6~T is 52.4\% lower than at 0.3~T.

This trend runs counter to the usual expectation that increasing field weakens pinning: the pinning energy typically decreases with field as the growing vortex density dilutes the pinning force available per vortex, increasing the creep rate $S \approx T/U$. Collective creep theory predicts that, in the elastic regime, the pinning energy goes as $U_{el}(B,J) = U_0(B)(J_c/J)^\mu \propto B^\nu J^{-\mu}$, where $U_0(B)$ is the field-dependent prefactor, $J$ is the local current density in the sample, for magnetic field $B$ and positive critical exponents $\nu$ and $\mu$ \cite{Abulafia1996, Cole2023}. From Fig.~\ref{fig:fig5}(d,e), discussed in detail later, we see no substantial change in $\mu$ with field at fixed temperature in the elastic regime that would point to a major change in the type of dynamics, e.g., increasing vortex bundle size or an elastic-to-plastic crossover, that could explain the slowing of creep with increasing vortex density. We therefore focus on the prefactor $U_0(B) \sim B^\nu$.

In the 3D small-bundle regime, $U_0(B)$ is well-approximated by $U_c$, the elastic energy cost of displacing a Larkin domain of volume $V_c = R_c^2 L_c$ by a distance of order $\xi_{ab}$, the displacement at which the lattice loses elastic correlation with the pinning landscape. This shear energy is $U_c \sim c_{66}(\xi_{ab}/R_c)^2 V_c$ \cite{Eley2018}, which simplifies to $U_c \approx c_{66}\xi_{ab}^2 L_c$, independent of $R_c$, where $c_{66}$ is the shear modulus of the vortex lattice \cite{Blatter1994}. The shear modulus itself depends on the vortex areal density $n_v$ as $c_{66} = (\Phi_0/8\pi\lambda_{ab})^2 n_v$, with $n_v = B/\Phi_0$, so that $c_{66} \propto B/\lambda_{ab}^2$ as the vortex density grows. Together with $L_c \propto a_0^{-1} \propto B^{1/2}$, where $a_0$ is the vortex spacing, this gives $U_c \propto B^{3/2}$, consistent with the positive exponent $\nu$ we extract. Physically, this reflects the stiffening of the vortex lattice with increasing field: a denser lattice is elastically harder to deform locally, raising the energy cost of a pinning-induced distortion and thereby suppressing creep, even as the growing vortex density might otherwise be expected to weaken pinning per vortex.

Now, focusing on the Fig. \ref{fig:fig5}(b) data at 15~K, we see that $S(H)$ is non-monotonic. The 0\% and 2.5\% Ca samples, for example, show a maximum in $S$ at 1~T, followed by minima $\mu_0H_{S,\min}$ near 3 and 4~T, respectively. Non-monononetic $S(H)$ has been observed in numerous superconductors whose irreversible magnetization loops exibit second magnetzation peaks, e.g. Ba(Fe$_{1-x}$Co$_x$)$_2$As$_2$ \cite{Prozorov2009}, (Ca$_{0.85}$La$_{0.15}$)$_{10}$(Pt$_3$As$_8$)(Fe$_2$As$_2$)$_5$ \cite{Sundar2023}, and HgBa$_2$CuO$_{4+\delta}$ \cite{Eley2020sr}. In such samples, Polichetti et al. \cite{Polichetti2021} reported a correlation between $\mu_0H_{S,\min}$ and the onset field $H_{\mathrm{onset}}$ of the second magnetization peak (SMP), a local minimum in the positive branch of $M(H)$ preceding the SMP. They found that $H_{\mathrm{onset}}$ consistently occurs below $H_{S,\min}$, with $H_{\mathrm{onset}}/H_{S,\min} < 1$ across several superconductors.

To test whether our films follow this trend, we compare $S(H)$ to the upper branch of the corresponding $m(H)$ loop in Fig.~\ref{fig:fig5}(c), marking $H_{\mathrm{smp}}$ and $H_{\mathrm{onset}}$. In both films, $S$ begins to decrease well below $H_{\mathrm{onset}}$, showing that creep suppression precedes the increase in magnetic moment. For the 0, 2.5, and 7.5\% Ca films (not shown in Fig.~\ref{fig:fig5}(c)), $H_{\mathrm{onset}}/H_{S,\min}$ is $0.80\pm0.13$, $0.74\pm0.12$, and $0.75\pm0.11$, respectively; at 10~K, the corresponding ratios for 0 and 2.5\% Ca are $0.67\pm0.08$ and $0.66\pm0.08$. These ratios are consistent with Polichetti et al., supporting a connection between the suppression of vortex creep and the development of the SMP. Additional $S(H)$ data at other temperatures are included in Supplementary Fig.~S9.

Next, we constructed the vortex phase diagram for 0 and 10\% Ca as shown in Figs.~\ref{fig:fig5}(e,f). To identify the elastic creep regime and estimate the vortex bundle size through the glassy exponent $\mu$, we define the experimentally accessible effective activation energy $U^* \equiv U_0+\mu k_BT\ln(t/t_0)= k_BT/S$ \cite{Zhou2016a, Sun_2015, Sun2015d,Haberkorn2011b,Miu2013,Sundar2017}. Combining Eq.~\ref{eq:ST} with the current-dependent activation barrier in Eq.~\ref{eq:Uinterpolation} and the Arrhenius relation $t = t_0 e^{U_{act}(J)/k_B T}$ is consistent with $U^* =U_0(J_{c0}/J)^\mu$.
Thus, $\mu$ can be extracted from the slope of $\ln (U^*)$ versus $\ln(1/J)$ which is plotted in Supplementary Fig.~S8. 

Using the extracted values of $\mu$, we identify distinct regions within the elastic creep regime in both films. For the film with 10\% Ca, a pronounced change occurs near $T/T_c\approx0.15$, where $\mu$ jumps from 1.44 to 0.61 at 0.5~T and from 1.61 to 0.76 at 1~T with increasing $T$. These values suggest a transition from creep involving small vortex bundles toward larger bundles, consistent with the collective creep framework \cite{Blatter1994, Abulafia1996}. This change in the characteristic vortex bundle size may contribute to the onset of the plateau in $S(T)$ near 18~K in Fig.~\ref{fig:fig4}(a). At 3~T, $\mu$ decreases from 1.88 to 1.18 with increasing temperature. The latter value is close to $\mu=1$, identified by Abulafia et al. as the prediction for intermediate-bundle creep within collective creep theory, suggesting a possible crossover toward this regime \cite{Abulafia1996}. At higher temperatures under 3~T, the slope of $\ln U^*$ versus $\ln(1/J)$ becomes negative, a behavior commonly associated with plastic creep \cite{Abulafia1996,Kierfeld2000, Eley2020sr,Burlachkov2022, Cole2023}.

We complete the diagrams by including the irreversibility field $H_{\mathrm{irr}}(T)$ and upper critical field $H_{c2}(T)$, with the intervening reversible mixed-state region identified as the vortex-liquid regime. To determine $H_{\mathrm{irr}}(T)$, we use fixed-field $M(T)$ sweeps, which provide a less noisy determination of the closure of the magnetic branches in our measurements than field-swept $M(H)$ loops. At each applied field, $T_{\mathrm{irr}}$ is defined as the temperature at which the upper and lower $M(T)$ branches merge within the measurement resolution. Details of the extraction of $H_{\mathrm{irr}}(T)$ are provided in Supplementary Fig.~S10. Similarly, $H_{c2}(T)$ is determined from fixed-field resistivity measurements, $\rho(T)$. At each applied field, $T_{c2}$ is defined using the criterion $\rho(T_{c2})=0.8\rho_{\mathrm{normal}}$, where $\rho_{\mathrm{normal}}$ is the normal-state resistivity. Details of the extraction of $H_{c2}(T)$ are provided in Supplementary Fig.~S1--S3.

The films with 0 and 10\% Ca exhibit qualitatively similar vortex phase diagrams. To highlight the primary changes induced by Ca substitution, we compare their characteristic field boundaries directly in Fig.~\ref{fig:fig5}(f). Ca substitution shifts the boundaries toward higher $T/T_c$. At $\mu_0H=3.5$~T, $T/T_c$ corresponding to the SMP onset increases from $T/T_c=0.063$ to 0.228. At 5~T, the irreversibility and upper critical field lines shift from $T/T_c=0.528$ to 0.646 and from 0.778 to 0.906, respectively. These shifts demonstrate that Ca substitution extends both the magnetically irreversible and superconducting regimes to higher $T/T_c$ at a given field, indicating improvements beyond the increase in $T_c$ alone.

\begin{figure}[!ht]
\centering
\includegraphics[width=1\linewidth]{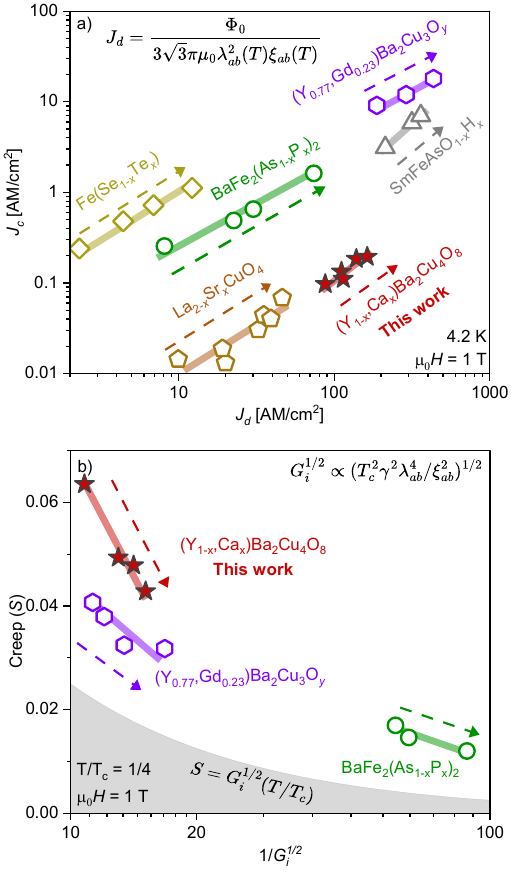}
\caption{\label{fig:fig6} (a) In-field critical current density $J_c(1\mathrm{\ T})$ as a function of the depairing current density $J_d$ at 4.2 K for Cu- and Fe-based superconductors \cite{Hodaka2021, Miura2022, Miura2024}. The $J_c$ values for YCa124 at 4.2 K are obtained by linear interpolation between the data measured at 1.8 and 5 K. (b) Vortex creep parameter $S$ as a function of $1/{G_i^{1/2}}$ at $T/T_c = 1/4$ and $\mu_0H = 1$ T for YGd123, YCa124, and Ba122:P superconductors \cite{Miura2022, Miura2024}. Where the shaded region denotes values of $S$ below the predicted lower limit \cite{Eley2017}.}
\end{figure}

Finally, we compare the current-carrying performance and vortex creep of YCa124 with those of other Cu- and Fe-based superconductors. Figure~\ref{fig:fig6}(a) shows that $J_c(\mathrm{1\ T})$ increases with $J_d$ across the Ca-doping series, following the positive correlation observed in other superconducting families \cite{Hodaka2021, Miura2022,Miura2024}. Between the undoped and 10\% Ca films, $J_d$ increases from approximately 88 to 163~MA/cm$^2$, accompanied by an increase in $J_c(\mathrm{1\ T})$ from approximately 0.097 to 0.194~MA/cm$^2$ at 4.2~K. Nevertheless, the measured $J_c(\mathrm{1\ T})$ remains well below $J_d$, indicating substantial scope for further improvement through inclusion of artificial vortex pinning centers. 

Figure~\ref{fig:fig6}(b) compares $S$ against $G_i^{-1/2}$ at $T/T_c=1/4$ and $\mu_0H=1$~T, with the gray region marking the predicted lower bound \cite{Eley2017}. Across the YCa124 series, $S$ spans approximately 0.043--0.064, while $G_i$ varies from approximately $4.4\times10^{-3}$ to $8.6\times10^{-3}$. The overall trend toward slower creep with decreasing $G_i$ is consistent with the behavior reported for carrier-tuned Cu- and Fe-based superconductors \cite{Miura2022,Miura2024}. Together, the increases in $J_d$ and $J_c$, accompanied by reduced creep, establish Ca substitution as an effective route for improving the superconducting performance of Y124.

\section*{Conclusions}

We investigated how Ca substitution modifies the microstructure, critical current density, and vortex dynamics of Y124 films with Ca concentrations up to 10\%. Electron microscopy revealed planar Y-125 intergrowths bounded by partial dislocation loops, consistent with the extended defects implicated by the pinning analysis. Ca substitution increases $J_c$ alongside the estimated depairing current density $J_d$, with the in-field $J_c$ at 4.2~K rising from approximately 0.097 to 0.194~MA/cm$^2$ between the undoped and 10\% Ca films. These improvements are accompanied by slower vortex creep and shifts of the irreversibility and upper critical field boundaries toward higher reduced temperatures. These results motivate extending the study to higher Ca concentrations to determine the limits of carrier-induced improvements in $J_c$ and creep suppression. Combining Ca substitution with artificial pinning centers offers a complementary route to further enhance performance by jointly tuning superconducting properties and the defect landscape.

\section*{Methods}\label{sec:Methods}
\subsection*{Film Growth}

The YCa124 films were grown by a molten hydroxide method. Yttrium oxide (Y$_2$O$_3$), calcium carbonate (CaCO$_3$), barium carbonate (BaCO$_3$), and copper oxide (CuO) powders served as the cation sources, combined in a nominal Y:Ca:Ba:Cu ratio of $(1-x)$:$x$:2:4 to target the composition (Y$_{1-x}$Ca$_x$)Ba$_2$Cu$_4$O$_8$; five Ca substitution levels were prepared ($x = 0$, 0.025, 0.05, 0.075, and 0.10). A total of 10 g of these precursor powders was combined with molten potassium hydroxide (KOH) solvent, which served as the hydroxide flux, and an NdGaO$_3$ (001) single-crystal substrate was submerged in the resulting mixture. Crystallization proceeded in a muffle furnace held at 650~\textdegree{}C in air for 12 h, following procedures first described in Ref.~\cite{Funaki2012}.

\subsection*{Magnetometry Measurements}

Magnetic characterization was performed using a Quantum Design MPMS3 superconducting quantum interference device (SQUID) magnetometer. All films were mounted in a standard MPMS3 brass holder. The magnetic field was applied parallel to the crystallographic \textit{c} axis, $H \parallel c$ (perpendicular to the film surface), and the magnetic moment was measured using a 10 mm scan length over a scan time of 1 s.

The superconducting transition temperature, $T_c$, was determined from measurements of the magnetic moment as a function of temperature, $m(T)$, acquired in an applied field of $\mu_0H = 0.3 \ \mathrm{mT}$. The temperature was swept at a rate of approximately 1.5 K/min.  The $T_c$ was defined as the temperature at which the difference between the normalized ZFC and FC magnetic moments first fell below $-0.02$. Finally, the magnetization, $M$, was calculated by dividing the measured magnetic moment by the sample volume. The sample dimensions are provided in Supplementary Table S1.

To characterize $J_c(T,H)$, we collected magnetic hysteresis loops $m(H)$ at fixed temperatures, while continuously sweeping the magnetic field at 100 Oe s$^{-1}$. The NdGaO$_3$ substrates contribute a noticeable paramagnetic background but exhibit no discernible magnetic hysteresis \cite{Eswara2023}. We therefore attribute the measured hysteresis to the superconducting films. At each temperature, the reversible background was estimated by averaging the upper and lower branches at the same applied field, $m_{\mathrm{bg}}(H)=[m_{\mathrm{upper}}(H)+m_{\mathrm{lower}}(H)]/2$, and subtracted from both branches to isolate the irreversible superconducting response.

Magnetic relaxation measurements were performed following conventional flux-creep protocols \cite{Yeshurun1996}. After establishing the critical state, the magnetic moment $m(t,H_1,T_1)$ was recorded every 4--5 s for approximately one hour at a fixed magnetic field $H_1$ and temperature $T_1$. The critical state was established by sweeping the magnetic field by $\Delta H =1 - 2\ \mathrm{T}$, corresponding to more than $4H^*$, where $H^*$ is the minimum field required for complete flux penetration. The field was then set to the desired value $H_1$ for the relaxation measurement. Establishment of the critical state was verified by comparing the initial relaxation measurement with the corresponding magnetic hysteresis loop, confirming that the initial point lay on the hysteresis curve.  

The measured moments were corrected for the background signal arising from the sample holder and substrate. The measurement time was then shifted to account for the interval between field stabilization and acquisition of the first data point, with the time offset chosen to maximize the correlation coefficient of the fit. The normalized creep rate, $S=-d\ln m/d\ln t$, was determined from a linear regression of $\ln m$ as a function of $\ln t$. Additional details regarding the relaxation-data analysis are provided in Fig.~S8 of the Supplementary Information.

\vspace{0.3cm}

\section*{Author Contributions}
S.E. and M.M. conceived and designed the experiment, and assisted with all data interpretation.
J.L. performed magnetization studies and associated data analysis.
B.Z. assisted with analysis of the magnetization data.
S.F. grew the (Y,Ca)Ba$_2$Cu$_4$O$_8$ films.
M.M., Y.O. and R.N. carried out the transport measurement. 
Y.O. analyzed the upper critical field data.
R.Y. and T.K. performed microstructural studies. 
J.L. and S.E. wrote the manuscript.
All authors commented on the manuscript.

\section*{Conflicts of Interest}

The authors declare no conflicts of interest.

\section*{Data availability}

The data supporting the findings of this study are available on Mendeley Data (DOI:10.17632/5tvtjh8nmz.1) as a zip file. This includes Origin file (.opju) that contain data spreadsheets for all the samples and figures used in this paper, which can be opened using Origin Viewer, a free application that permits viewing and copying of data contained in Origin project files. 

\section*{Acknowledgments}

This work was supported by the National Science Foundation through the University of Washington Materials Research Science and Engineering Center under Grant No. DMR-2308979 (J.L. and S.E.) and in part by NSF Grant No. DMR-2046925 (S.E.).
Work performed at Seikei University was supported by JST FOREST (Grant No. JPMJFR202G, Japan) and in part by JSPS KAKENHI (Grant Nos. 26K00986 and 23KK0073).

\title{Supplementary information for: Enhanced Critical Currents and Irreversibility Fields in YBa$_2$Cu$_4$O$_8$ Films through Ca-Substitution}

\date{}

\author{%
Jiangteng Liu,$^{1}$ Shuhei Funaki,$^2$ Yuki Ogimoto,$^3$ Bryan Zhang,$^1$ Ryoya Nagaura,$^{3}$ Ryuji Yoshida,$^{3,4}$ Takeharu Kato,$^3$ Masashi Miura$^{3,5}$, Serena Eley$^{1*}$\\
$^1$Department of Electrical \& Computer Engineering, University of Washington, Seattle, WA 98195 USA\\
$^2$Faculty of Engineering, Aichi Institute of technology: 1247 Yachigusa Yakusa-chou, Toyota-shi, Aichi 470-0392, Japan\\
$^3$Graduate School of Science and Technology, Seikei University: 3-3-1 Kichijoji-kitamachi, Musashino-shi, Tokyo 180-8633, Japan\\
$^4$Nanostructures Research Laboratory, Japan Fine Ceramics Center:   2-4-1 Mutsuno, Atuta-ku, Nagoya 456-8587, Japan\\
$^5$JST-FOREST, 7, Gobancho, Chiyoda-ku, Tokyo 102-0076, Japan\\
\underline{$^{1}$serename@uw.edu}\\}

\pagebreak
\let\clearpage\relax 
\onecolumngrid
\section*{Supplemental Materials}\label{sec:smaterials}
% \title{Supplemental Materials}
\setcounter{figure}{0}

\makeatletter 
\renewcommand{\thefigure}{S\@arabic\c@figure}
\makeatother

\setcounter{figure}{0}
\makeatletter 
\renewcommand{\thefigure}{S\@arabic\c@figure}
\makeatother

\setcounter{table}{0}
\makeatletter 
\renewcommand{\thetable}{S\@arabic\c@table}
\makeatother

\newcommand{\vect}[1]{\mathbf{#1}}
% \sisetup{load-configurations = abbreviations}
\DeclareSIUnit\oersted{Oe}

\maketitle

\begin{figure*}[h!]
\centering
\includegraphics[width=1\linewidth]{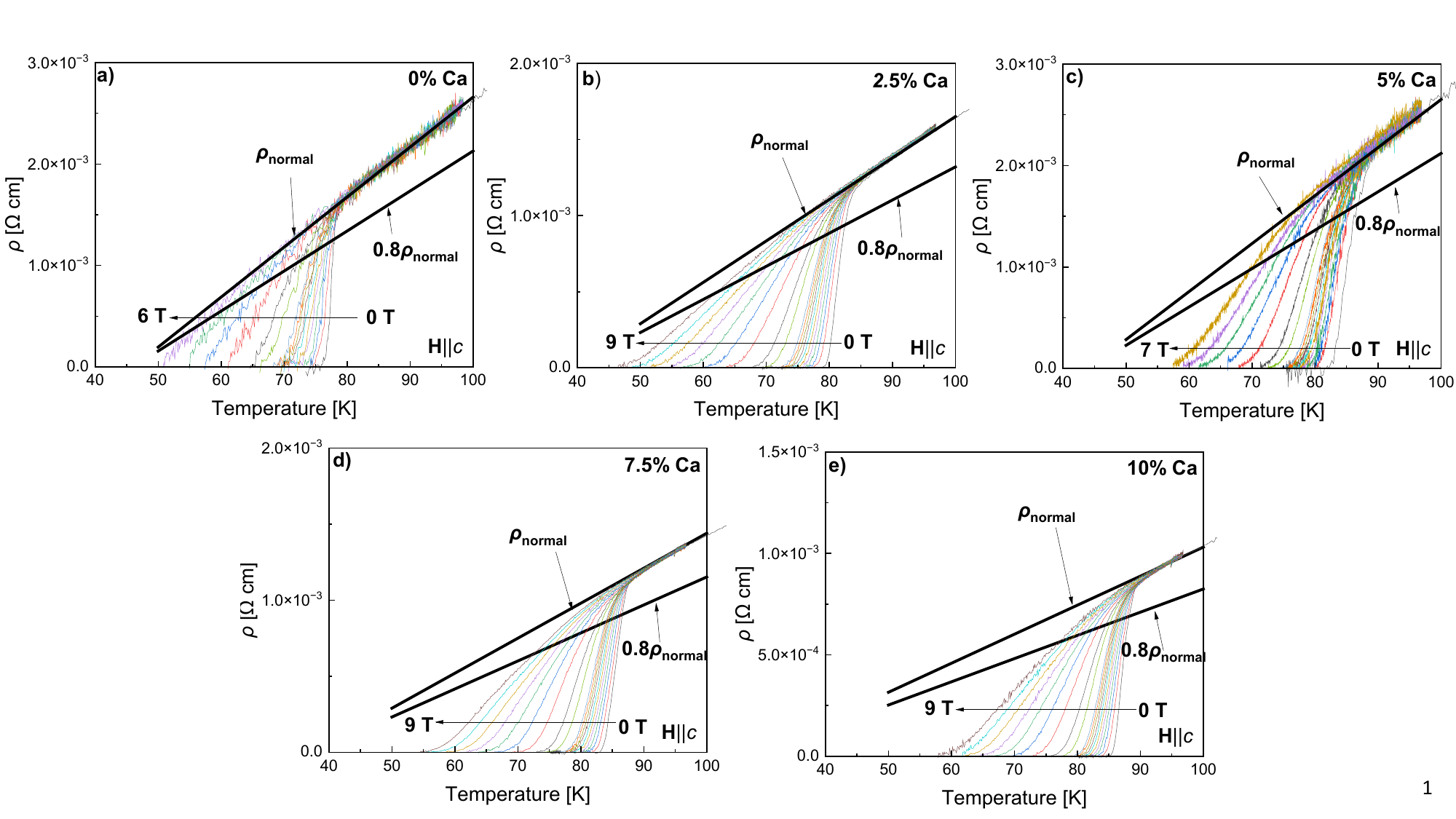}
\caption{\label{fig:figS1} Temperature dependence of the resistivity $\rho(T)$, measured at different fixed magnetic fields applied parallel to the $c$ axis ($H\parallel c$) for YCa124 films with (a) 0\% Ca, (b) 2.5\% Ca, (c) 5\% Ca, (d) 7.5\% Ca, and (e) 10\% Ca. At each applied field, $T_{c2}(H)$ is defined by the criterion $\rho (T_{c2})=0.8\rho_{\mathrm{normal}}$, where $\rho_{\mathrm{normal}}$ is obtained by extrapolating a linear fit of the normal-state resistance into the superconducting transition region. }
\end{figure*}

\begin{figure*}[h!]
\centering
\includegraphics[width=1\linewidth]{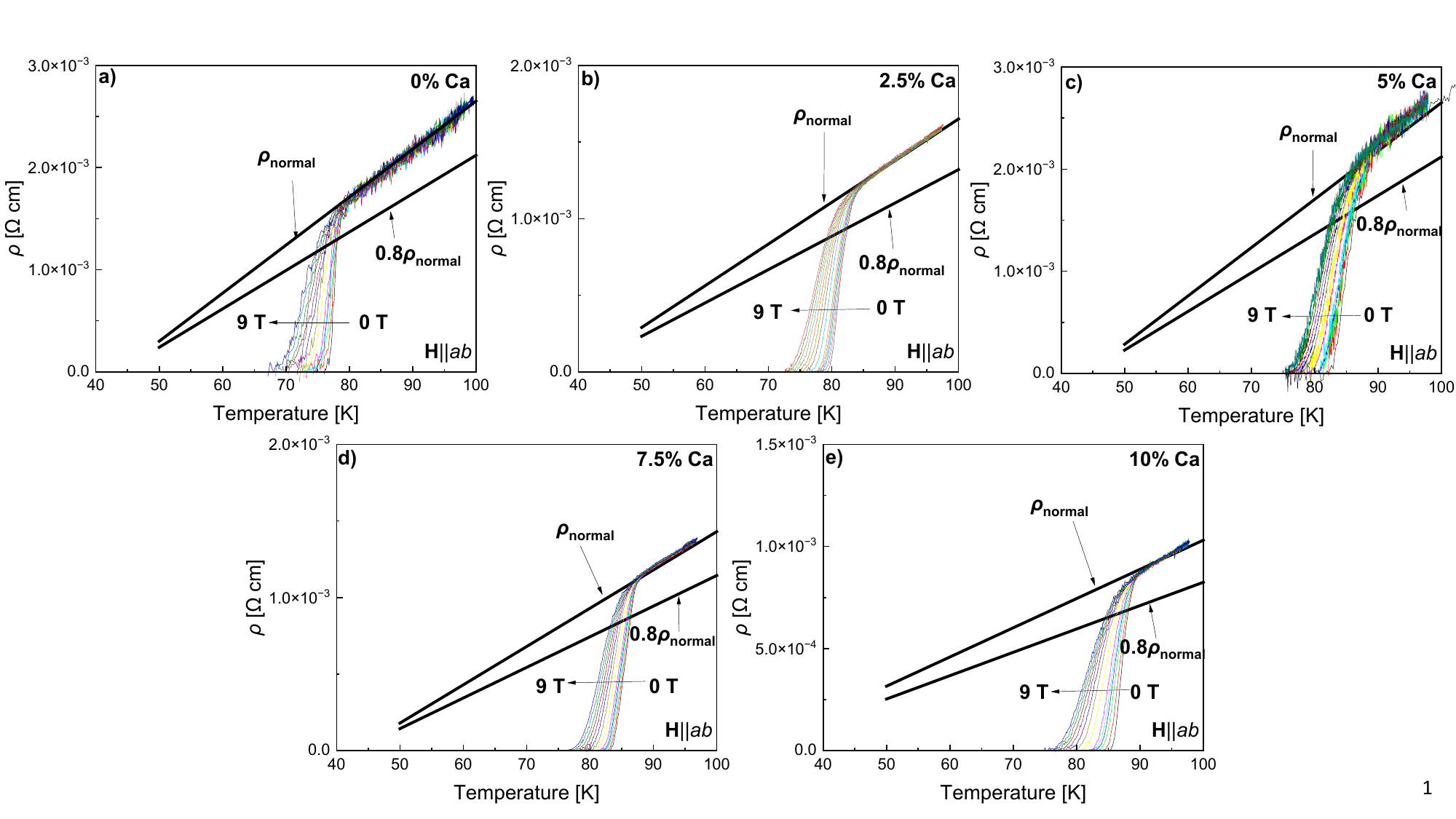}
\caption{\label{fig:figS2} Temperature dependence of the resistivity $\rho(T)$, measured at different fixed magnetic fields applied parallel to the $ab$ axis ($H\parallel ab$) for YCa124 films with (a) 0\% Ca, (b) 2.5\% Ca, (c) 5\% Ca, (d) 7.5\% Ca, and (e) 10\% Ca. At each applied field, $T_{c2}(H)$ is defined by the criterion $\rho(T_{c2})=0.8\rho_{\mathrm{normal}}$, where $\rho_{\mathrm{normal}}$ is obtained by extrapolating a linear fit of the normal-state resistance into the superconducting transition region. }
\end{figure*}

\begin{figure*}[h!]
\centering
\includegraphics[width=1\linewidth]{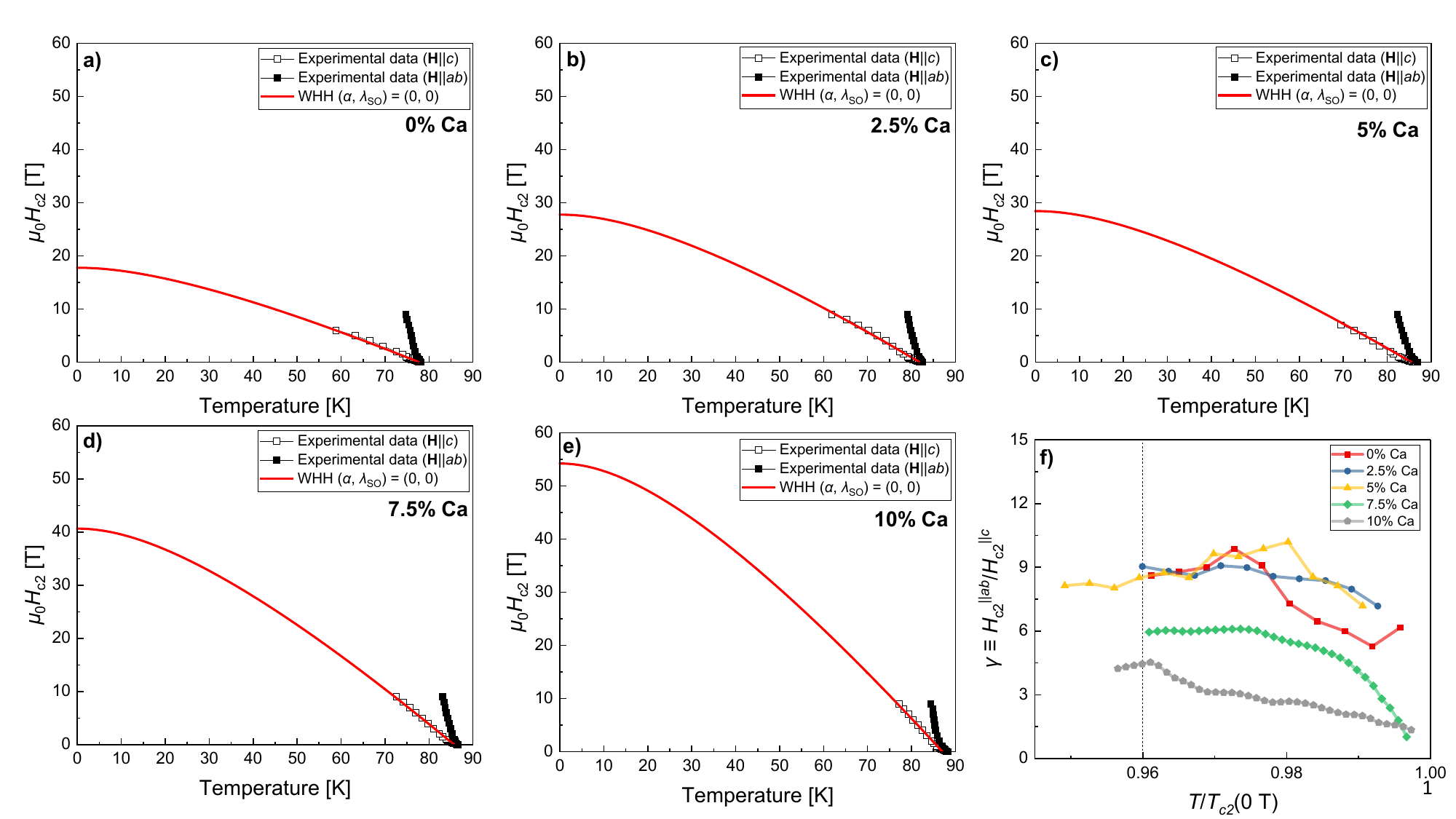}
\caption{\label{fig:figS3 }Temperature dependence of the upper critical field $H_{c2}(T)$, extracted from resistivity measurements for magnetic fields applied parallel to the $c$ axis ($H\parallel c$) and the $ab$ plane ($H\parallel ab$) for YCa124 films with (a) 0\% Ca, (b) 2.5\% Ca, (c) 5\% Ca, (d) 7.5\% Ca, and (e) 10\% Ca. The red line represents fits to the Werthamer–Helfand–Hohenberg (WHH) model with the Maki parameter $\alpha=0$ and spin–orbit scattering parameter $\lambda_{\mathrm{SO}}=0$. (f) Temperature dependence of the upper critical field anisotropy $\gamma\equiv H_{c2}^{\parallel ab}/H_{c2}^{\parallel c}$, for all five YCa124 films.}
\end{figure*}

\begin{figure*}[h!]
\centering
\includegraphics[width=1\linewidth]{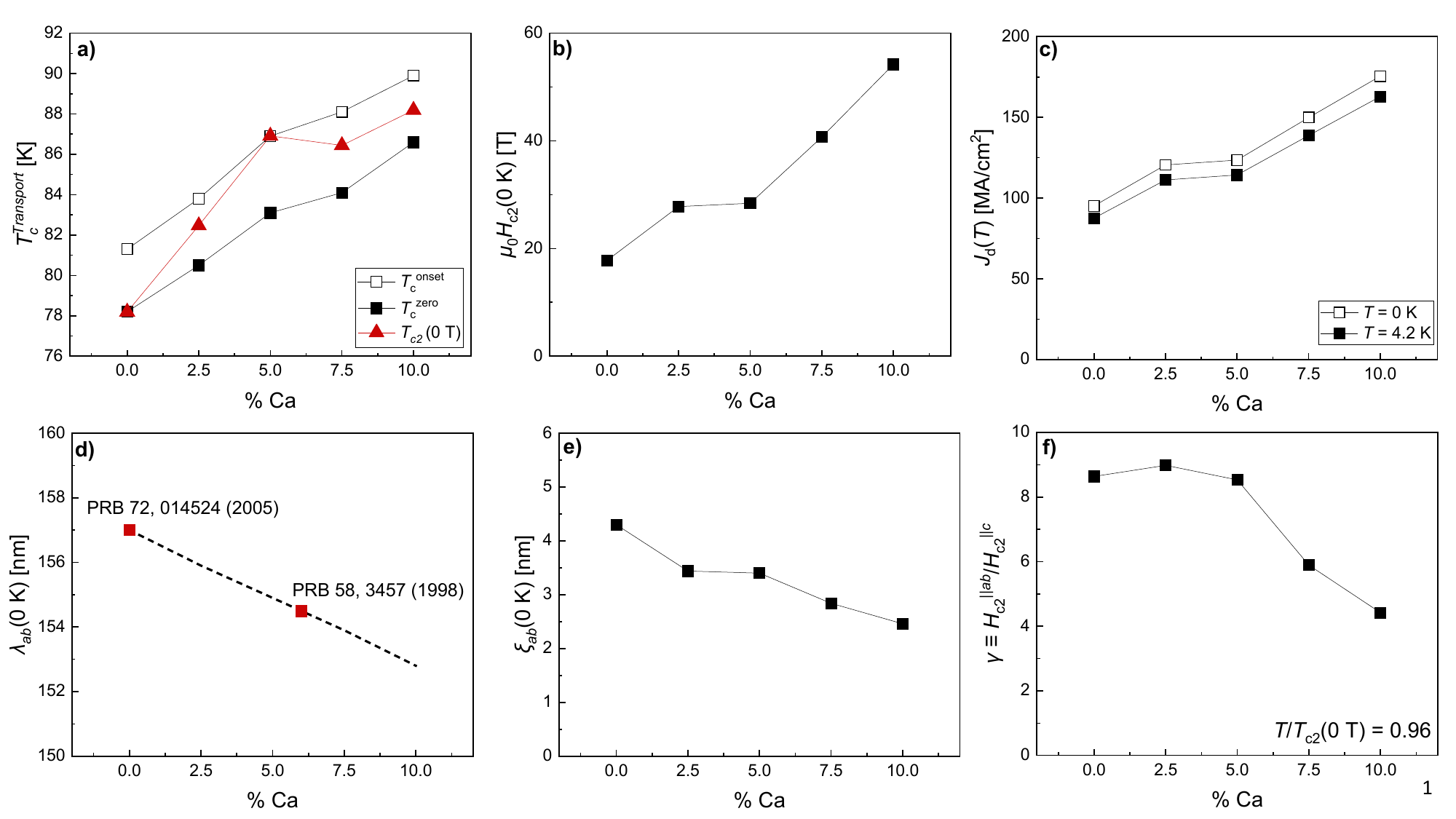}
\caption{\label{fig:figS4} Summary of superconducting parameters for all five YCa124 films. (a) Onset and offset transition temperatures and $T_{c2}(0\,\mathrm{T})$. (b) Upper critical field $H_{c2}^{\parallel c}$, extrapolated to zero temperature using the WHH model. (c) Calculated $J_c$ at 0 and 4.2~K. (d) Zero-temperature in-plane penetration depth $\lambda_{ab}$ obtained from the literature. (e) Zero-temperature in-plane coherence length calculated using $\xi_{ab}=[\Phi_0/(2\pi\mu_0H_{c2}^{\parallel c})]^{1/2}$, where $\Phi_0$ is the magnetic flux quantum. (f) Upper critical field anisotropy $\gamma\equiv H_{c2}^{\parallel ab}/H_{c2}^{\parallel c}$, evaluated at $T/T_{c2}(0\,\mathrm{T})=0.96$. }
\end{figure*}

\begin{figure*}[h!]
\centering
\includegraphics[width=0.5\linewidth]{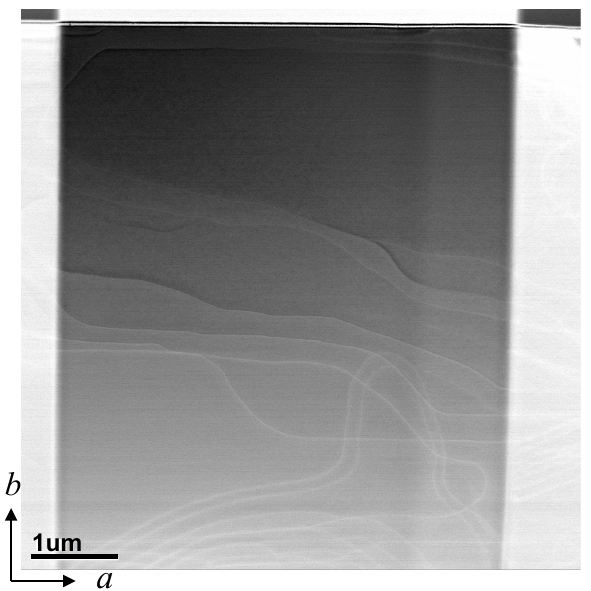}
\caption{\label{fig:figS5} Planar-view STEM dark field image showing the stacking fault for the film with 7.5\% Ca. }
\end{figure*}

\begin{figure*}[h!]
\centering
\includegraphics[width=1\linewidth]{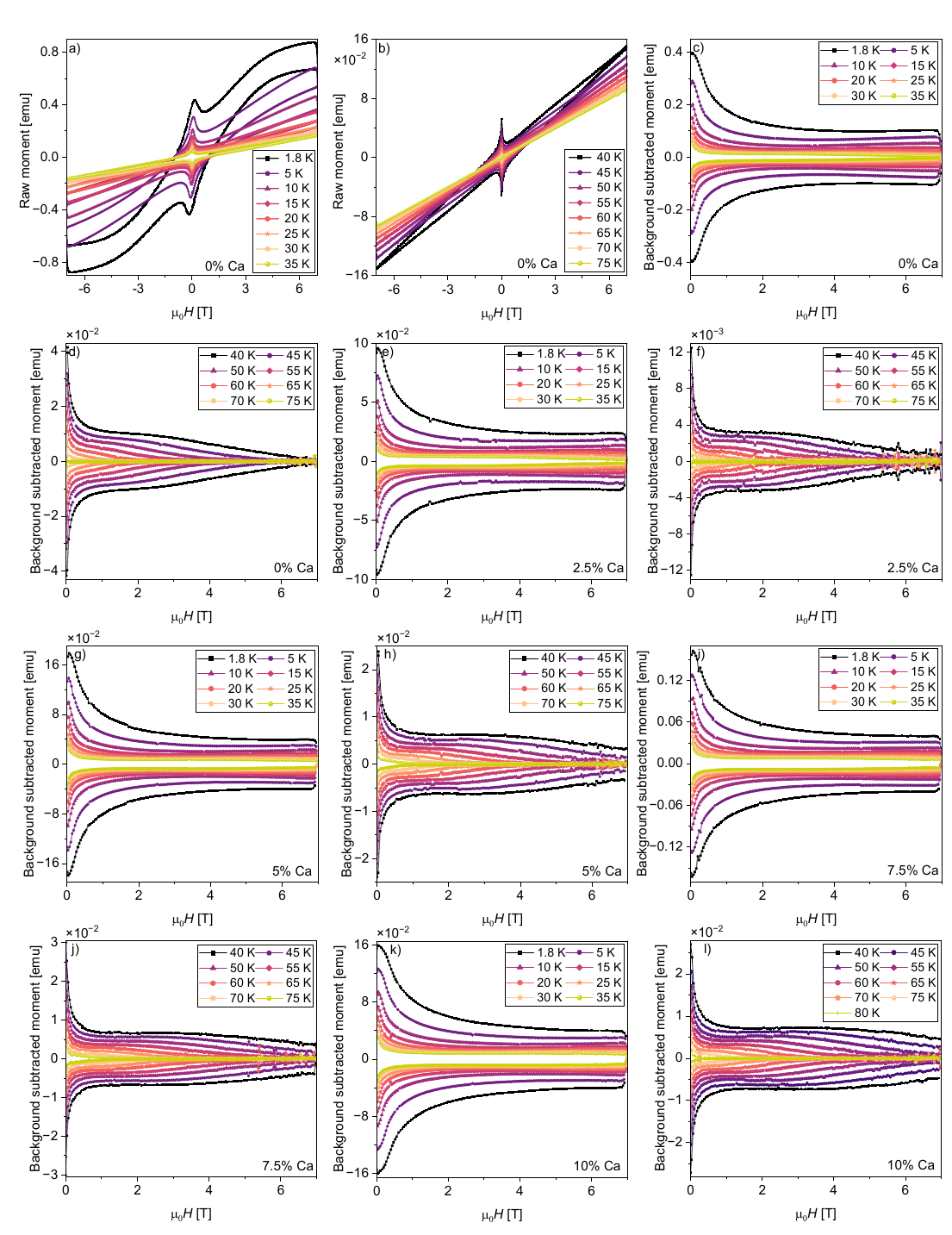}
\caption{\label{fig:figS6} (a,b) Example raw magnetic hysteresis loops for the 0\% Ca film, showing the background contribution from the NdGaO$_3$ substrate. Background-subtracted magnetic hysteresis loops measured over the 0--7~T field range at different temperatures for (c,d) 0\% Ca, (e,f) 2.5\% Ca, (g,h) 5\% Ca, (i,j) 7.5\% Ca, and (k,l) 10\% Ca films.}
\end{figure*}

\section*{Magnetic Relaxation}

\begin{figure*}[h!]
\centering
\includegraphics[width=1\linewidth]{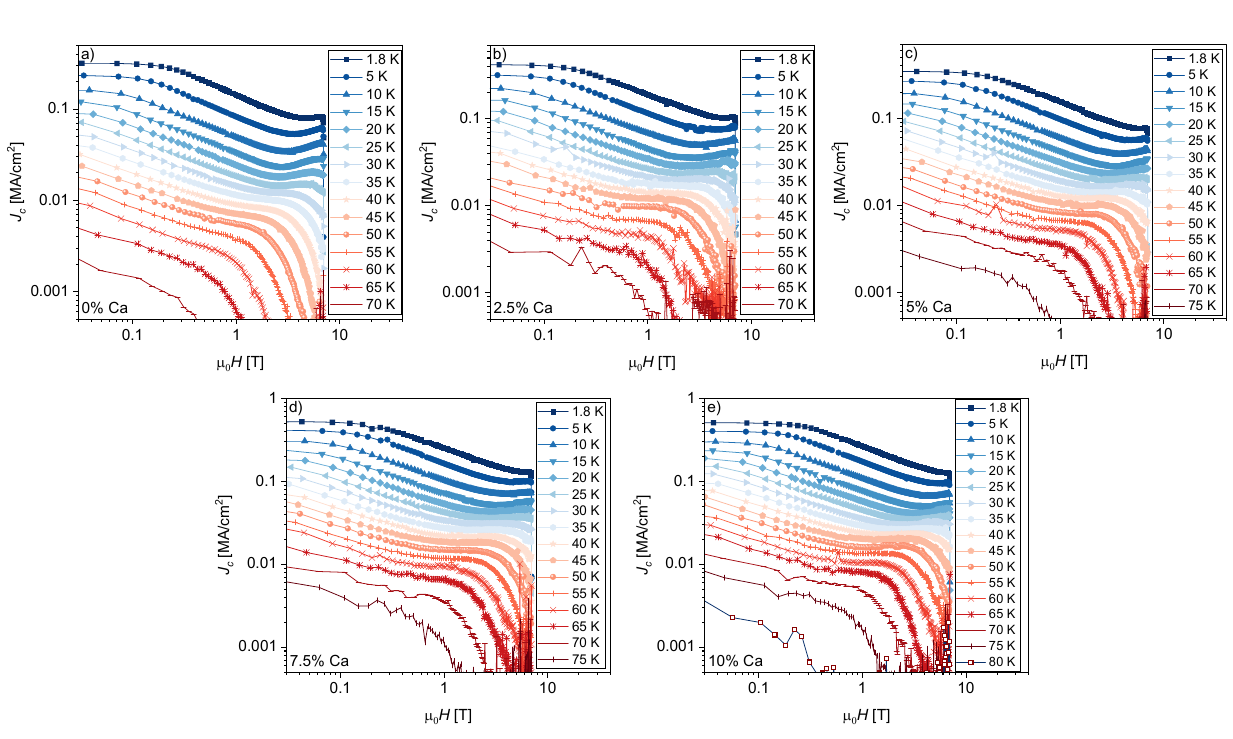}
\caption{\label{fig:figS7} Field dependence of the critical current density $J_c$ at various temperatures for YCa124 films with (a) 0\% Ca, (b) 2.5\% Ca, (c) 5\% Ca, (d) 7.5\% Ca, and (e) 10\% Ca. }
\end{figure*}

\begin{figure*}[h!]
\centering
\includegraphics[width=1\linewidth]{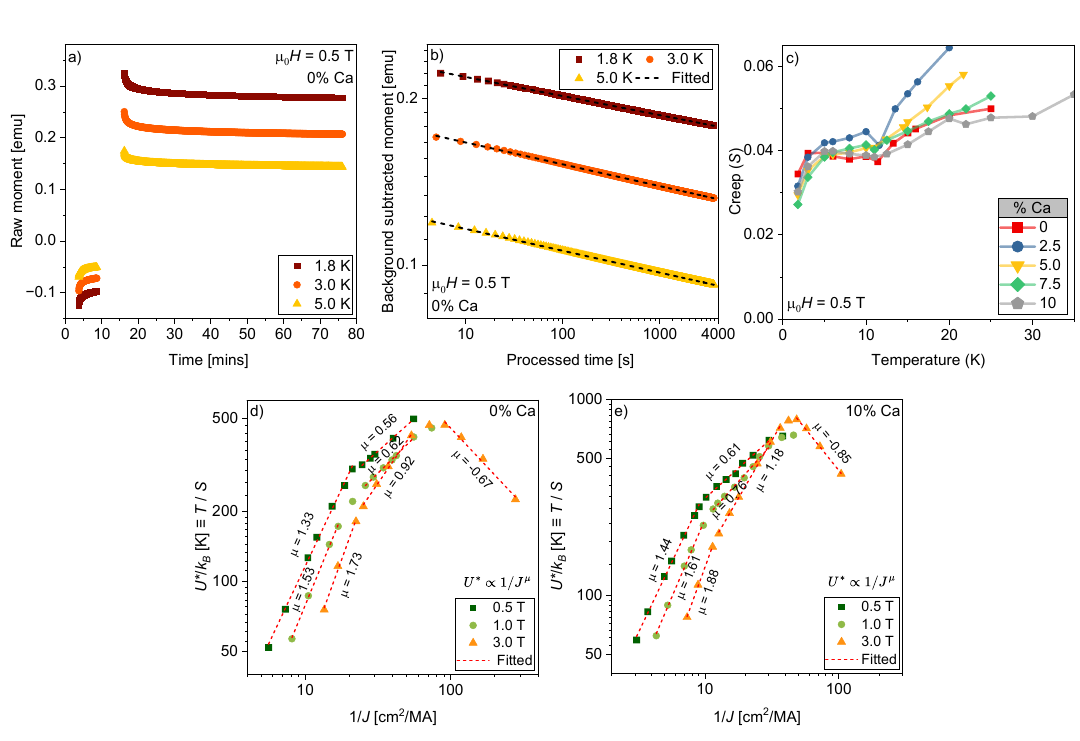}
\caption{\label{fig:figS8} 
(a) Representative raw magnetic relaxation data, $m(t)$, for the 0\% Ca film measured at $\mu_0H = 0.5$~T at various temperatures. (b) Corresponding relaxation curves after accounting for the system delay time $t_d$, plotted as $\ln m$ versus $\ln(t+t_d)$. The vortex creep parameter $S$ is obtained from the slope of the linear fit. (c) Temperature dependence of the relaxation rate $S$ at $\mu_0H = 0.5$~T for YCa124 films with different Ca concentrations. Effective activation energy $U^*/k_B \equiv T/S$ as a function of $1/J$ at different applied magnetic fields for YCa124 films with (d) 0\% Ca and (e) 10\% Ca.}
\end{figure*}

Figure~\ref{fig:figS8}(a) shows the example raw magnetic relaxation data for 0\% Ca film measured under an applied field of $\mu_0 H = 0.5$~T. Relaxation was recorded for 5 minutes on the lower branch and for 60 minutes on the upper branch of the hysteresis loop, in both cases after preparing the sample in the critical state. The background contribution from the sample mount and NdGaO$_3$ substrate was estimated by averaging the sum of the magnetic moments measured on the upper and lower branches and dividing by two. The resulting background signal was then subtracted from the upper branch relaxation data. Because the MPMS3 acquires the first data point after a short delay, during which relaxation has already begun. Therefore, we introduce a system delay time $t_d$ when extracting the vortex creep rate. The creep parameter is defined as $S = -d \ln m/d \ln (t + t_d)$,
where $t_d$ is treated as a fitting parameter chosen to maximize the linear correlation. The fully processed moment data used to obtain $S$ are shown in Fig.~\ref{fig:figS8}(b). Figure~\ref{fig:figS8}(c) presents additional results for the temperature dependence of $S$ at $\mu_0H = 0.5$~T for all YCa124 samples. Figures~\ref{fig:figS8}(d,e) show the effective activation energy, $U^*/k_B \equiv T/S$, as a function of $1/J$ at different applied magnetic fields for the 0 and 10\% Ca films. The slopes of these plots are used to determine the glassy exponent $\mu$, and the resulting $\mu$ values are used to identify the elastic-creep regime shown in the vortex phase diagrams in Figs.~5(d,e) of the main text. 

\begin{figure*}[h!]
\centering
\includegraphics[width=1\linewidth]{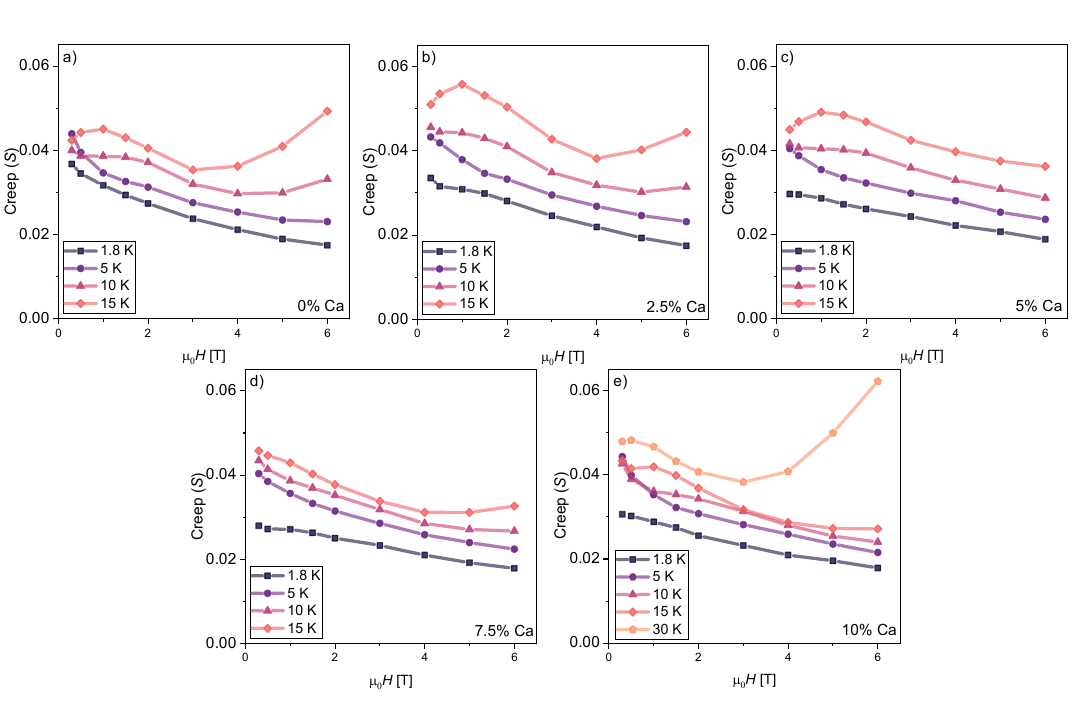}
\caption{\label{fig:figS9} Field dependence of the vortex creep rate $S$ at different temperatures for YCa124 films with (a) 0\% Ca, (b) 2.5\% Ca, (c) 5\% Ca, (d) 7.5\% Ca, and (e) 10\% Ca.}
\end{figure*}

\section*{Irreversibility Line}

The irreversibility line, $H_{\mathrm{irr}}(T)$, was determined from temperature-dependent magnetization measurements performed at fixed magnetic fields after establishing the Bean critical state. At each field, the separation between the upper and lower magnetization branches was calculated as
\[
\Delta m(T)=\left|m_{\mathrm{lower}}(T)-m_{\mathrm{upper}}(T)\right|.
\]
To quantify the background branch separation and measurement noise, the mean value, $\overline{\Delta m}_{\mathrm{bg}}$, and standard deviation, $\sigma_{\mathrm{bg}}$, of $\Delta m$ were calculated over the temperature range 90--98~K, which is above $T_c$ for all samples. The irreversibility temperature, $T_{\mathrm{irr}}$, was defined as the first measured temperature, upon increasing temperature, at which the branch separation fell below the criterion
\[
\Delta m(T) <
\overline{\Delta m}_{\mathrm{bg}}+50\sigma_{\mathrm{bg}}.
\]
The uncertainty in $T_{\mathrm{irr}}$ was estimated using corresponding criteria of
\[
\overline{\Delta m}_{\mathrm{bg}}+25\sigma_{\mathrm{bg}}
\quad \mathrm{and} \quad
\overline{\Delta m}_{\mathrm{bg}}+75\sigma_{\mathrm{bg}}.
\]
Because $\Delta m$ decreases with increasing temperature as the irreversible magnetization vanishes, the $75\sigma_{\mathrm{bg}}$ criterion gives the lower-temperature bound of $T_{\mathrm{irr}}$, whereas the $25\sigma_{\mathrm{bg}}$ criterion gives the upper-temperature bound. The resulting $T_{\mathrm{irr}}$ values at each applied field were used to construct the $H_{\mathrm{irr}}(T)$ lines shown in the main text. Representative fixed-field $M(T)$ measurements used for this analysis are shown in Fig.~S10.

\begin{figure*}[h!]
\centering
\includegraphics[width=1\linewidth]{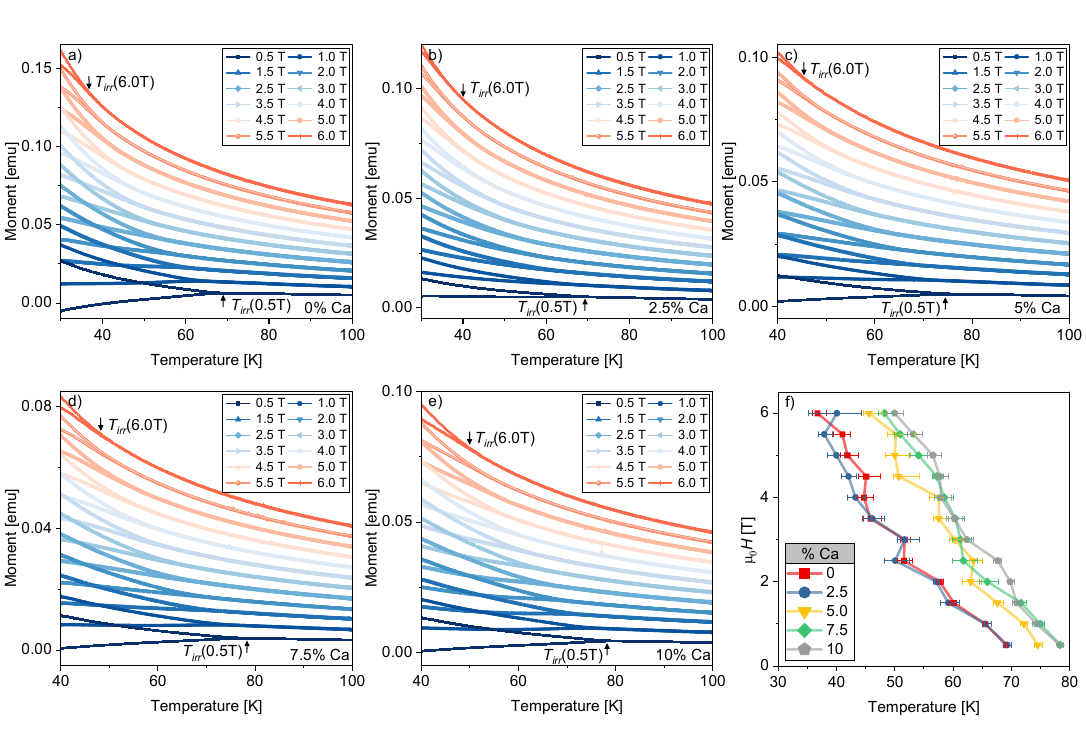}
\caption{\label{fig:figS10} Temperature sweeps at fixed magnetic fields for the upper and lower branches after establishing the Bean critical state for YCa124 films with (a) 0\% Ca, (b) 2.5\% Ca, (c) 5\% Ca, (d) 7.5\% Ca, and (e) 10\% Ca. The irreversibility temperature, $T_{\mathrm{irr}}(\mu_0H)$, is determined from the temperature at which the separation between the upper and lower branches falls below the threshold defined in the text. The field-dependent magnetic moment observed above $T_{\mathrm{irr}}$ originates from the background contribution of the NdGaO$_3$ substrate.}
\end{figure*}

\begin{table}[h!]
\caption{Sample dimensions}
\label{tab:samples1}
\begin{center}
\begin{tabular*}{0.9\textwidth}{@{\extracolsep{\fill}}ccccccc}
\hline\hline
sample & Ca concentration & length $l$ & width $w$ & thickness $\delta$\\
\# & \% & mm & mm & $\mu$m\\
\hline
1 & 0 & $2.755\pm0.083$ & $2.344\pm0.129$ & $45.884\pm1.849$\\
2 & 2.5 & $2.355\pm0.170$ & $1.824\pm0.070$ & $15.724\pm1.481$\\
3 & 5 & $2.490\pm0.097$ & $1.837\pm0.042$ & $32.589\pm2.414$\\
4 & 7.5 & $2.110\pm0.075$ & $1.764\pm0.026$ & $26.202\pm2.967$\\
5 & 10 & $2.388\pm0.049$ & $1.709\pm0.106$ & $23.597\pm2.102$\\
\hline\hline
\end{tabular*}
\end{center}
\end{table}

\end{document}